\documentclass[aps,prc,twocolumn,superscriptaddress,amsmath,amssymb]{revtex4-2}

\usepackage{graphicx}% Include figure files
\usepackage{dcolumn}% Align table columns on decimal point
\usepackage{bm}% bold math
\usepackage{amsmath, amssymb}
\usepackage{threeparttable, multirow, booktabs, longtable}
\usepackage{color}

\begin{document}
\setlength{\LTcapwidth}{\textwidth}

%\preprint{}

\title{Shell-model study on properties of proton dripline nuclides with $Z, N$ = 30-50 including uncertainty analysis}


%\author{}
%\thanks{}
%\affiliation{}

\author{Boshuai Cai}
\affiliation{Sino-French Institute of Nuclear Engineering and Technology, Sun Yat-Sen University, Zhuhai, 519082, Guangdong, China}
\author{Guangshang Chen}
\affiliation{Sino-French Institute of Nuclear Engineering and Technology, Sun Yat-Sen University, Zhuhai, 519082, Guangdong, China}
\author{Cenxi Yuan}
\email{yuancx@mail.sysu.edu.cn}
\affiliation{Sino-French Institute of Nuclear Engineering and Technology, Sun Yat-Sen University, Zhuhai, 519082, Guangdong, China}
\author{Jianjun He}
\email{hejianjun@bnu.edu.cn}
\affiliation{Key Laboratory of Beam Technology and Materials Modification of Ministry of Education, College of Nuclear Science and Technology, Beijing Normal University, Beijing 100875, China}
\affiliation{Beijing Radiation Center, Beijing 100875, China}


\date{\today}

\begin{abstract}

	The binding energies and proton separation energies of nuclides with $Z, N = 30-50$ are investigated, based on the shell model with an uncertainty analysis through statistical methods. Several formulas are used to obtain the binding energies and proton separation energies according to the shell-model calculations. The non-parametric Bootstrap method is applied to establish an uncertainty decomposition and recomposition framework. Moreover, it is used to estimate the stability of proton(s) emission for each nuclide. Two formulas for calculating the binding energies with a systematic uncertainty of $\sim0.3$ MeV are proposed, and a reliable extrapolation ability is examined. These binding energy formulas deduce similar forms of respective $S_{p}$ and $S_{2p}$ energies, which predict the extension of the nuclear boundary of this region. A nice description of the binding energies and proton separation energies is provided. The one- and two-proton separation energies and partial half-lives of proton emitting are predicted, thus showing a new dripline. Besides, there are 30 unstable nuclides predicted to be bound against proton(s)-emission. These nuclear properties will be useful in nuclear astrophysics.
	

\end{abstract}

%\keywords{}

\maketitle

%\section{}
%\subsection{}
%\subsubsection{}

% If in two-column mode, this environment will change to single-column
% format so that long equations can be displayed. Use
%\begin{widetext}
%\end{widetext}

\section{Introduction}\label{section1}
	
	The accurate description of the synthesis of the heavy elements in the Universe is one of the main open questions in the field of nuclear astrophysics. The type of particles involved in nuclear process is one of the foundation to investigate it \cite{bertulani2020, liu2020network}. While the majority of heavy elements are produced in two neutron-induced processes, known as slow ($s$) and rapid ($r$) processes \cite{kappeler2011s, arnould2007r}, there are 35 neutron-deficient stable nuclei (from $^{74}$Se to $^{196}$Hg), the so-called $p$ nuclei, which cannot be created in these scenarios but related to the $p$-process (or $\gamma$-process) with photodisintegration reactions of ($\gamma$, $n$), ($\gamma$, $p$) and ($\gamma$, $\alpha$) \cite{arnould2003p}. Apart from the nucleosynthesis in nature, there are roughly 7000 possible candidates on the nuclear landscape. However,  up to now, only roughly 3000 nuclides have been identified and a vast territory remains to be explored \cite{thoennessen2011isotopes, erler2012limits, nazarewicz2018limits}. The location of the dripline, defined unambiguously by the nucleon or two-nucleon separation energies \cite{thoennessen2004reaching}, depicts the boundary of the nuclear landscape, which is essential for understanding the relationships between the total number of nuclides and the nuclear force \cite{thoennessen2016discovery}. 
	
	The masses and half-lives of nuclei are important inputs for the nuclear astrophysics. Accurate measurements for nuclei are thus of significant importance in nuclear astrophysics. Unfortunately, the measurement of unknown nuclear masses is not always feasible even possible with the up to date experimental technology. Experimentally, only the neutron dripline of nuclides with $Z\leqslant10$ are determined \cite{ahn2019location}. Thus, theoretical models become crucial in order to evaluate and predict the unknown masses.

	Various mass models have been proposed and corrected in the last several decades. Myers and Swiatecki developed the semi-empirical droplet model of nuclear masses and deformations \cite{myers1966nuclear,myers1976development}. The finite-range droplet model, which was developed from a Yukawa-plus-exponential macroscopic model (finite-range liquid-drop model) and a folded-Yukawa single-particle potential for the microscopic energy, are proposed by $\rm{M\ddot{o}ller}$ \emph{et al.} in 1981 and corrected in the following years \cite{moller1981atomic, moller1981nuclear, moller1988nuclear, moller1988nuclear2, moller1993nuclear, moller2016nuclear}. In addition, the Hartree-Fock-Bogoliubov was applied to the determination of nuclear masses, and many versions have been developed with the Skyrme-type and the Gogny-type effective interactions \cite{goriely2013hartree, wang2014surface, ge2019effect}. The time-dependent Hartree-Fock was used to describe the multinucleon transfer dynamics \cite{wu2020production}.
	
	In this paper, we carry out a shell-model study on proton dripline properties of nuclides with $Z, N=30-50$,\ which are of astrophysical $\nu$p-process importance.  In addition, some nuclei are also important in the Type I X-ray bursts \cite{schatz2017dependence, parikh2009impact}. A full $f_{5}pg_{9}$ shell model calculation is applied to investigate the binding energies, proton separation energies, and the proton dripline properties (including the partial half-lives of proton emitting), for these exotic nuclei. The formulas of energies calculation are introduced in Sec. \ref{sec:models}.  Furthermore, the confidence of our results is also examined with an uncertainty decomposition framework \cite{cai2020alpha} based on the Bootstrap method \cite{efron1982jackknife, efron1992bootstrap}, of which the framework is presented in Sec. \ref{sec:framework}. Results are discussed in Sec. \ref{sec:results}.

\section{\label{section2}Formulas and statistic method}
	
\subsection{\label{sec:models} Calculation of binding energy and separation energies}

	In recent years, some properties of light proton-rich nuclei  were well studied in experiment, while the shell model provides reasonable theoretical descriptions, e.g., the isospin asymmetry \cite{lee2020isospin, jian2021isospin}, the $\beta$-delayed proton(s) emission \cite{xu2017Si22, liang2020P26}, the $\beta$-decay spectroscopy \cite{sun2019S27, wu2020Al22}, the exotic $\beta$-$\gamma$-$\alpha$ decay mode of $^{20}$Na \cite{wang2021Na20}, the four-proton unbound nucleus $^{18}$Mg \cite{jin2021Mg18}, etc. In neutron-deficient heavy nuclei, the shell model well described the $\alpha$-decay of the lightest isotope of U \cite{zhang2021U214} and isomeric states of $^{218}$Pa \cite{zhang2020Pa218} and $^{213}$Th \cite{zhou2021Th213}. For medium mass nuclei, investigations locate mostly in neutron-rich nuclei \cite{yuan2016Sn132,chen2019Sn132, watanabe2021Ag127}  and neutron-deficient nuclei with $Z\leqslant N$ \cite{yun2020what, xu2019In101, jin2021Cd98} while rarely exceeds beyond $Z=N$.

	The shell model provides lots of descriptions for nuclear spectroscopic properties, but rarely investigates the binding energy. A reasonable choice of effective interaction is key to perform the shell-model calculation. Through the interaction, shell-model calculations provide the valence part of  binding energy without considering the Coulomb interaction. It is noted as $E_{BE, SM}(Z, N)$, where $Z$ $(N)$ denotes the proton (neutron) number. The effective interaction JUN45, which was proposed by Honma \emph{et al.} via fitting to the experimental data of selected nuclides in this $f_{5}pg_{9}$ shell model space consisting of four single-particle orbits $p_{3/2}$, $f_{5/2}$, $p_{1/2}$ and $g_{9/2}$ \cite{honma2009new}, is used in the present work for calculating $E_{BE,SM}(Z,N)$. Necessarily, to estimate the binding energy based on the shell model, several corrections should be taken into account. In 1997, Herndl and Brown \cite{herndl1997shell}  used an overall constant ($cst$) and four terms linear and quadratic in the number of valence protons and that of valence neutrons:
\begin{equation}
\begin{split}
E_{BE}(Z,N)=E_{BE,SM}(Z,N)+cst+a(Z-28)\\
+b(Z-28)^2+c(N-28)^2+d(N-28),
\label{eq:herndl}
\end{split}
\end{equation}
where $a$, $b$, $c$ and $d$ are fitting parameters. This form is equivalent to the combination of two body Coulomb interaction and small variation in the nuclear size and mass  \cite{herndl1997shell}. 

	As an analog of Eq. (\ref{eq:herndl}), one can pose a quadratic formula for the binding energy of a certain nuclide with fixing the constant term of the original formula by the binding energy of $^{56}$Ni:	
	\begin{equation}
	\begin{split}
		E_{BE}(Z,N)=E_{BE, SM}(Z, N) + E_{BE}(^{56}\text{Ni}) + a(Z-28)\\ + b(Z-28)^{2}
		 + c(N-28)^{2} + d(N-28).
		\label{eq:1}
	\end{split}
	\end{equation}
	The correction of Coulomb energy is included in the $a$ and $b$ terms. In addition, the single-particle energies of the core, which are kept unchanged for all nuclides, may actually depend on the numbers of valence nucleons and should be compensated through introducing additional terms on $Z$ and $N$, as suggested in Eq. (\ref{eq:1}). Through fitting results, we suggest to replace the last term of Eq. (\ref{eq:1}) by the difference between valence protons and valence neutrons:
	\begin{equation}
	\begin{split}
		E_{BE}(Z,N)=E_{BE, SM}(Z,N) + E_{BE}(^{56}\text{Ni}) + a(Z-28) \\+ b(Z-28)^{2} 
		 + c(N-28)^{2} + d(Z-N)^{2}.
		\label{eq:yuan}
	\end{split}
	\end{equation}
Besides, the sum of the last two terms and the residuals of Eqs. (\ref{eq:1}) and (\ref{eq:yuan}) shows a hyperboloid-like distribution, it is thus approached with a term $(Z-30)(Z-2N+50)$:
	\begin{equation}
	\begin{split}
		E_{BE}(Z,N)=E_{BE, SM}(Z,N) + E_{BE}(^{56}\text{Ni}) + a(Z-28) \\+ b(Z-28)^{2} 
		+ c(Z-30)(Z-2N+50).
		\label{eq:chen}
	\end{split}
	\end{equation}
Furthermore, Caurier $et \ al.$ \cite{caurier1999full} proposed a correction including modified Coulomb energy term and a monopole expression:
	\begin{equation}
	E_{BE}(Z,N)=E_{BE, SM}(Z,N) + E_{BE}(^{56}\text{Ni}) + E_{C} + E_{M},
	\label{eq:E.Caurier}
	\end{equation}
	with
	\begin{equation}
	\nonumber
	\begin{split}
	E_{C} = a(Z-28) + b(Z-28)(Z-29) \\
	 + c(Z-28)(N-28),	\\
	E_{M} = d(A-56) + e(A-56)(A-57) \\
	 + f(T(T+1)-\frac{3}{4}(A-56)),	
	\end{split}
	\end{equation}
	where $T$ is the isospin of nuclide and $A$ is the mass number. Comparing Eq. (\ref{eq:yuan}) and Eq. (\ref{eq:E.Caurier}), they are quite similar. The difference is a term involving the gap of total isospin. While focusing on the ground state property, a general rule shows that $T \approx T_{3} \propto |N-Z|$. Thus the term involving isospin could also be absorbed in the quadratic form of nucleon numbers.
	
	Nucleon separation energies can be easily calculated through the binding energy with formulas $S_{p}(Z, N)=E_{BE}(Z,N)-E_{BE}(Z-1,N)$ and $S_{2p}(Z, N)=E_{BE}(Z,N)-E_{BE}(Z-2,N)$. From both Eqs. (\ref{eq:yuan}) and (\ref{eq:chen}), general forms of $S_{p}$ and $S_{2p}$ are deduced:
	\begin{equation}
	\begin{aligned}
		S_{p}(Z, N) &= E_{BE, SM}(Z, N) - E_{BE, SM}(Z-1, N) \\
		& \quad + aZ + bN + d,
		\label{eq:sp}
	\end{aligned}
	\end{equation}
	\begin{equation}
	\begin{aligned}
		S_{2p}(Z, N) &= E_{BE, SM}(Z, N) - E_{BE, SM}(Z-2, N) \\
		& \quad + aZ + bN + d.
		\label{eq:s2p}
	\end{aligned}
	\end{equation}
	
	Notwithstanding the quadratic term, $c(Z-N)^{2}$, disappears by subtracting two binding energies, it shows correlation with the residuals of Eqs. (\ref{eq:sp}) and (\ref{eq:s2p}). After reintroducing this term,
	\begin{equation}
	\begin{split}
		S_{p}(Z, N)=E_{BE, SM}(Z, N) - E_{BE, SM}(Z-1, N) \\
		 + aZ + bN + c(Z-N)^{2} + d,
		\label{eq:sp02}
	\end{split}
	\end{equation}
	\begin{equation}
	\begin{split}
		S_{2p}(Z, N)=E_{BE, SM}(Z, N) - E_{BE, SM}(Z-2, N) \\
		 + aZ + bN + c(Z-N)^{2} + d.
		\label{eq:s2p02}
	\end{split}
	\end{equation}
	perform well on describing the experimental data with lower uncertainties listed in TABLE \ref{tab:stat}. Furthermore, the value (uncertainty) of the fitting parameter $b$ is much decreased (increased) after the reintroduction of $c(Z-N)^{2}$, which shows unnecessity and unreliability of $b$. The term $bN$ is no more robust in Eqs. (\ref{eq:sp02}) and (\ref{eq:s2p02}) and will increase the statistical uncertainty of formulas. As expected, removing this term makes the following formulas perform better:
	\begin{equation}
	\begin{split}
		S_{p}(Z, N)=E_{BE, SM}(Z, N) - E_{BE, SM}(Z-1, N) \\
		 + aZ + c(Z-N)^{2} + d,
		\label{eq:sp03}
	\end{split}
	\end{equation}
	\begin{equation}
	\begin{split}
		S_{2p}(Z, N)=E_{BE, SM}(Z, N) - E_{BE, SM}(Z-2, N)  \\
		 + aZ + c(Z-N)^{2} + d.
		\label{eq:s2p03}
	\end{split}
	\end{equation}
	
	The performance of these formulas for binding energies and separation energies is investigated by an application of the Bootstrap method, which is introduced in the following section.

\subsection{\label{sec:framework} Framework of uncertainty analysis}

	The total uncertainty of a formula is composed of the experimental uncertainty, the statistical uncertainty and the systematic uncertainty \cite{dobaczewski2014, yuan2016uncertainty}. Our previous work \cite{cai2020alpha} detailed the practical steps of the Bootstrap method \cite{efron1982jackknife, efron1992bootstrap} for decomposing and recomposing uncertainties. Recently, Jia \emph{et al.} used this to study the correlation between parameters of Woods-Saxon potential and evaluate the uncertainty of the binary cluster model \cite{jia2021possible}. The benefit of such a method is that all useful statistics could be obtained simultaneously from the parameter space estimated from the resampling of the dataset. Note that the experimental uncertainties of binding energies and separation energies are neglected here because they are generally (more than 95\%) smaller than 0.1 MeV. The residual between theoretical and experimental data of a nuclide with $Z$ protons and $N$ neutrons is defined as,
	\begin{equation}
		r(Z, N, S_{BS, i}) = y_{cal}(Z, N, S_{BS, i})-y_{exp}(Z, N),
		\label{eq:rsd}
	\end{equation}
	where $y$ denotes the corresponding energy, subscript $cal$ denotes the calculated value, subscript $exp$ denotes the experimental value, and $S_{BS, i}$ denotes the $i$th Bootstrap sample in the $M$ Bootstrap samples, which are the same size as the original dataset. In practice, the number of the Bootstrap samples is set to be $M = 10^6$, which assures a robust estimation of uncertainties. The ordinary least squares method is used to make the fitting. Under this definition, the positive value of $r(Z, N, S_{BS, i})$ indicates an overestimation, while the negative one indicates an underestimation. The statistical uncertainty of a formula for a nuclide is defined by the unbiased standard deviation
	\begin{equation}
		\hat{\sigma}_{stat}^{2}\left(Z, N\right)=\frac{1}{M-1} \sum_{i=1}^{M}\left(y_{cal}\left(Z, N, S_{BS, i}\right) - \bar{y}_{cal}\left(Z, N\right)\right)^{2},
		\label{eq:sigma_stat_single}
	\end{equation}
	where $\bar{y}_{cal}(Z, N)$ is the mean of calculated values of a given nuclide. The global statistical uncertainty of a formula is the root-mean-square (rms) of $\hat{\sigma}_{\operatorname{stat}}\left(Z, N\right)$:
	\begin{equation}
		\hat{\sigma}_{stat}^{2}=\frac{1}{K} \sum_{k=1}^{K} \hat{\sigma}^{2}_{stat}\left(Z, N\right),
		\label{eq:sigma_stat_all}
	\end{equation}
	where $K$ is the number of combinations of protons and neutrons in the dataset.
	
	By ignoring the experimental uncertainty, the systematic uncertainty, which yields the gaps between the calculated and the ``true" values, is estimated by
	\begin{equation}
		\begin{aligned}
		\hat{\sigma}^{2}_{sys}\left(Z, N\right)&=\left(\sum_{i=1}^{M} \frac{y_{cal}\left(Z, N, S_{BS, i}\right)}{M} - y_{exp}\left(Z, N\right)\right) ^{2}	\\
		&= \left( \frac{1}{M} \sum_{i=1}^{M} r\left(Z, N, S_{BS, i}\right) \right)^{2},
		\end{aligned}
		\label{eq:sigma_sys_single}
	\end{equation}
	and the global systematic uncertainty is derived by:
	\begin{equation}
		\hat{\sigma}_{sys}^{2}=\frac{1}{K} \sum_{k=1}^{K} \hat{\sigma}^{2}_{sys}\left(Z, N\right) = \frac{1}{K} \sum_{k=1}^{K} \bar{r}^{2}\left(Z, N\right),
		\label{eq:sigma_sys_all}
	\end{equation}
	The total uncertainty of a property for a nuclide is defined by the rms of residuals
	\begin{equation}
		\begin{aligned}
		\hat{\sigma}_{total}^{2}\left(Z, N\right)  & =\frac{1}{M} \sum_{i=1}^{M} r^{2}\left(Z, N, S_{BS, i}\right) \\
		& =\frac{M-1}{M} \hat{\sigma}_{stat}^{2}\left(Z, N\right)+\hat{\sigma}_{sys}^{2}\left(Z, N\right),
		\end{aligned}
		\label{sigma_total_all}
	\end{equation}
	and then generalized to the whole dataset:
	\begin{equation}
		\hat{\sigma}_{total}^{2} = \hat{\sigma}_{sys}^{2} + \hat{\sigma}_{stat}^{2}.
		\label{sigma_relationship}
	\end{equation}
	In addition, the total uncertainty is recomposed as 
	\begin{equation}
		\begin{aligned}
		\hat{\sigma}_{pred}^{2}\left(Z, N\right)  = \hat{\sigma}_{stat}^{2}\left(Z, N\right)+\hat{\sigma}_{sys}^{2},
		\end{aligned}
		\label{eq:sigma_pred_all}
	\end{equation}
	to check the extrapolation power.
	
\begin{figure}
\includegraphics[width=.237\textwidth]{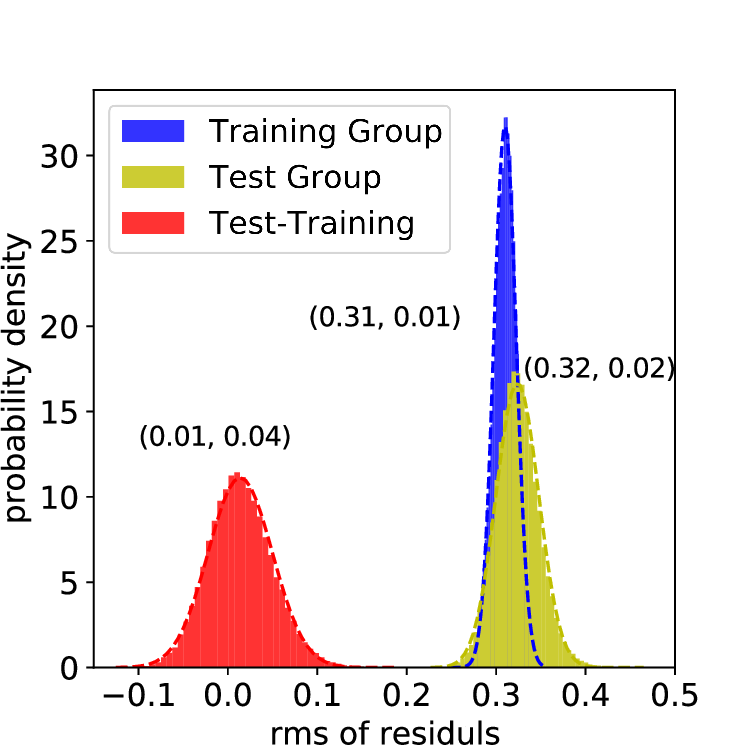}
\includegraphics[width=.237\textwidth]{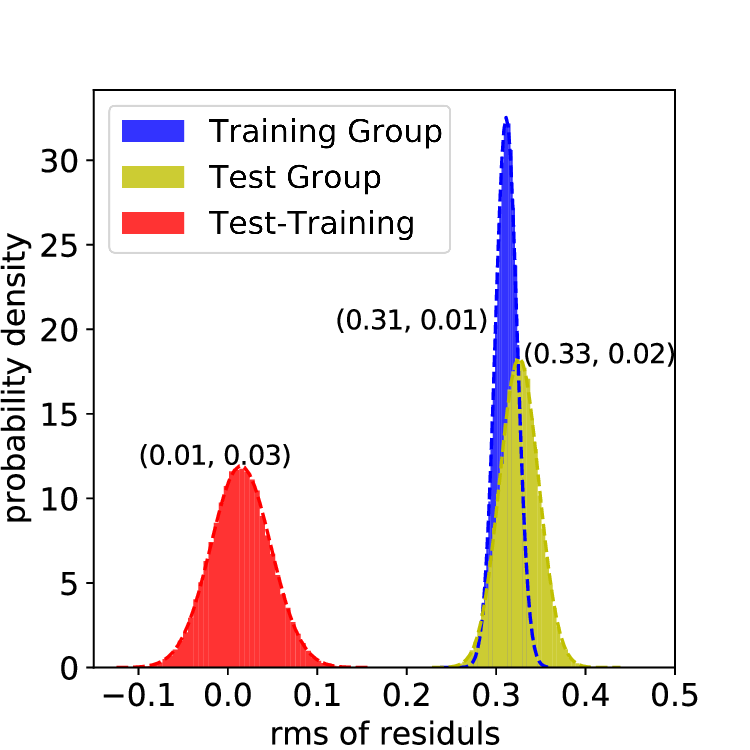}\\
\includegraphics[width=.237\textwidth]{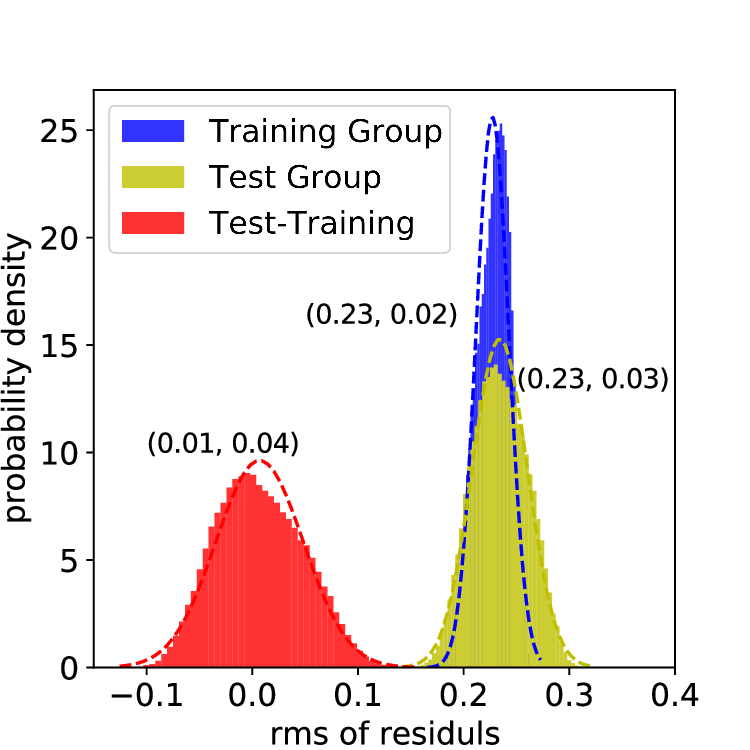}
\includegraphics[width=.237\textwidth]{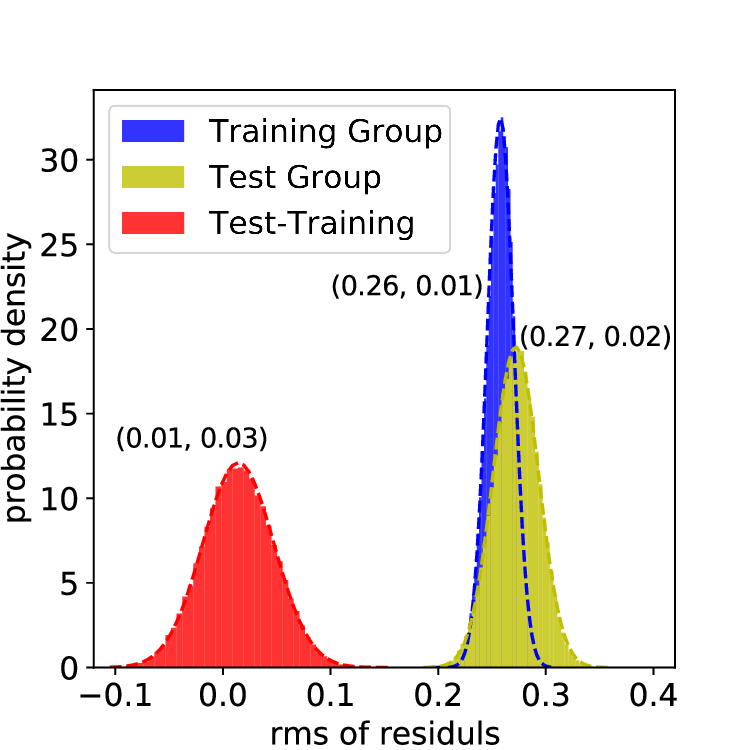}
\caption{\label{fig:st-ts}Distribution of uncertainties of training group, test group and their difference. The top two are for Eq. (\ref{eq:yuan}) and Eq. (\ref{eq:chen}). The bottom two are for Eq. (\ref{eq:sp04}) and Eq. (\ref{eq:s2p03}). The dash lines are the fitted normal distribution, of which the parameters, mean and standard deviation, are listed in the nearby parentheses.}
\end{figure}

	In effect, each time when a Bootstrap sample is obtained, the original dataset is divided to two parts: 1) the training group, in which nuclides form the Bootstrap sample; 2) the test group consisting of nuclides not included in the Bootstrap sample. For each Bootstrap sample, the uncertainty of the training group is 
\begin{equation}
\sigma_{tr}(S_{BS,i})=\sqrt{\frac{1}{n_{i,tr}}\sum_{k=1}^{n_{i,tr}}r^2(Z_k,N_k,S_{BS,i})},
\label{eq:tr}
\end{equation}
and that of the test group is 
\begin{equation}
\sigma_{ts}(S_{BS,i})=\sqrt{\frac{1}{n_{i,ts}}\sum_{k=1}^{n_{i,ts}}r^2(Z_k,N_k,S_{BS,i})},
\label{eq:ts}
\end{equation}
where $n_{i,tr}$ ($n_{i,ts}$) is the number of nuclides in the training (test) group. In other words, $\sigma_{tr}$ and $\sigma_{ts}$ describe the systematic uncertainties of random interpolation and extrapolation. The distribution of the uncertainties of the training and test groups are drawn in FIG. \ref{fig:st-ts} for the binding energy formulas Eqs. (\ref{eq:yuan}) and (\ref{eq:chen}) and the separation energy formulas Eqs. (\ref{eq:sp04}) and (\ref{eq:s2p03}). The $\sigma_{tr}$ locates within the distribution of $\sigma_{ts}$. Their difference $\sigma_{tr}-\sigma_{ts}$ locates around 0.01. $\sigma_{ts}$ distributes a little wider than $\sigma_{tr}$ but not statistically significant. The robustness of the systematic uncertainty estimation can be seen through this check. Note that the statistical uncertainty estimation is restricted to the sub-parameter-space of the correction terms because the uncertainty of JUN45 interaction is not significant as discussed later.

	As for the prediction of stabilities of $p$-emission and $2p$-emission, i.e., $P(S_{p}(Z, N)<0)$ and $P(S_{2p}(Z, N)<0)$, one could integrate them directly by assuming normalized distributions of $S_{p}(Z, N)$ energy and $S_{2p}(Z, N)$ energy which are established from the parameters space of formula and the corresponding systematic uncertainty.
	
	The results of the application on the binding energy formulas Eqs. (\ref{eq:1})-(\ref{eq:E.Caurier}) and the separation energy formulas Eqs. (\ref{eq:sp})-(\ref{eq:s2p03}) are presented and discussed in the following section.

\section{\label{sec:results}Results and discussions}

	In this work, experimental data are filtrated from AME2016 \cite{wang2017ame2016} within the nuclear landscape $Z, N \in [30, 50]$, corresponding to the $f_{5}pg_{9}$ shell. 221 nuclides with experimentally determined (values without \#) binding energy are obtained to perform the preliminary computation. As for $S_{p}$ and $S_{2p}$, regions are narrowed respectively to $Z, N\in [31, 50]$ and $Z, N\in [32, 50]$, including 198 nuclides with experimentally determined $S_{p}$ and 178 nuclides with experimentally determined $S_{2p}$. The choice is to avoid uncertainty introduced by calculation of nuclides with $Z, N$ = 28 and 29. Besides these measured, there are in principle 220 nuclides for $E_{BE}$, 202 nuclides for S$_{p}$, and 183 nuclides for S$_{2p}$ to be predicted. In practice, it is not necessary to predict all unknown nuclides in the model space because many of them are far beyond the proton dripline. 

\begin{table*}
\caption{\label{tab:stat}The fitting parameters, decomposed uncertainties for experimentally determined nuclides and the recomposed uncertainties for extrapolation of binding energy, $S_{p}$ and $S_{2p}$ formulas with standard deviation, unit is MeV.}
\begin{center} 
\footnotesize \tabcolsep 2pt 
\begin{threeparttable}
\begin{tabular*}{\textwidth}{c|ccc|cc|cccccc}\toprule[0.65pt] 
& $\sigma_{total}$\tnote{*} & $\sigma_{stat}$\tnote{*} & $\sigma_{sys}$\tnote{*} & $\sigma_{total}$\tnote{\#} & $\sigma_{stat}$\tnote{\#} & $a$  & $b$ & $c$ & $d$ & $e$ & $f$ \\
\hline
$E_{BE}(Eq.\ref{eq:1})$ & 0.731& 0.122&0.721			   &0.774&0.281& 			-9.15 (5) & -0.0954 (25) & 0.0188 (12) & -0.276 (31) & --- & --- \\
$E_{BE}(Eq.\ref{eq:yuan})$ 		&0.316&0.0482&0.312&0.416&0.275&			-9.61 (1) & -0.0944 (8) & 0.0245 (5) & -0.0262 (9) & --- & --- \\
%$Eq.\ref{eq:yuan}$\tnote{m}	&0.317&0.0849&0.305&	---&		---&			-9.62 (3) 	&-0.0858 (37)	&0.0175 (18)&	-0.0108 (44) & &\\
%$Eq.\ref{eq:yuan}$\tnote{o}	&0.254&0.0473&0.250&	---&		---&			-9.53	 (2)	&-0.0968 (8)	&0.0229 (5)&	-0.0250 (9)&	&\\
%$Eq.\ref{eq:yuan}$\tnote{a2m}	&0.379&0.0437&0.377&	---&		---&					---&			---&		---&			---&	---& ---       \\
%$Eq.\ref{eq:yuan}$\tnote{a2o}	&0.286&0.0499&0.281&	---&		---&					---&			---&		---&			---&	---& ---       \\
%$Eq.\ref{eq:yuan}$\tnote{m2o}	&0.867&0.305&0.812&	---&		---&					---&			---&		---&			---&	---& ---       \\
%$Eq.\ref{eq:yuan}$\tnote{o2m}	&0.557&0.0569&0.554&	---&		---&					---&			---&		---&			---&	---& ---       \\
$E_{BE}(Eq.\ref{eq:chen})$ &0.317&0.0399&0.315&0.374&0.202&			-9.12 (1) & -0.0921 (7) & -0.0309 (5) & --- & ---	& --- \\
%$Eq.\ref{eq:chen}$\tnote{m} 	&0.327&0.0768&0.317&	---&		---&			-9.19 (2)	&-0.0873 (21)	&-0.0276 (13) & --- & ---	& --- \\
%$Eq.\ref{eq:chen}$\tnote{o} 	&0.295&0.0426&0.292&	---&		---&			-9.10 (1)	&-0.0932 (7)	&-0.0305 (7)& --- & ---	& --- \\
%$Eq.\ref{eq:chen}$\tnote{a2m} 	&0.355&0.0431&0.352&	---&		---&					---&			---&		---&			---&	---& ---       \\
%$Eq.\ref{eq:chen}$\tnote{a2o} 	&0.300&0.0385&0.298&	---&		---&					---&			---&		---&			---&	---& ---       \\
%$Eq.\ref{eq:chen}$\tnote{m2o} 	&0.461&0.153 &0.434 &	---&		---&					---&			---&		---&			---&	---& ---       \\
%$Eq.\ref{eq:chen}$\tnote{o2m} 	&0.382&0.0724&0.375&	---&		---&					---&			---&		---&			---&	---& ---       \\
$E_{BE}(Eq.\ref{eq:E.Caurier})$ &0.305&0.0563&0.300&0.779&0.719& 	-9.45 (8) & -0.118 (1) & 0.425 (98) & 0.0221 (158) & -0.0939 (244) & 0.363 (96) \\
 \hline
$S_{p}(Eq.\ref{eq:sp})$ &0.286&0.0362&0.284&0.294&0.0781& 		-0.242 (6) & 0.0548 (52) & --- &-4.28 (19)	& --- & --- \\
$S_{p}(Eq.\ref{eq:sp02})$ &0.276&0.0404&0.273&0.375&0.257& 		-0.186 (16) & 0.00385 (1541) & 0.00335 (88) & -4.36 (19) & --- & --- \\
$S_{p}(Eq.\ref{eq:sp03})$ &0.275&0.0331&0.273&0.278&0.0519&	 	-0.182 (5) & --- & 0.00359 (27) & -4.36 (19) & --- & --- \\
$S_{p}(Eq.\ref{eq:sp04})$ &0.231&0.0328&0.228&0.233&0.0447&	 	-0.183 (4) & --- & 0.00364 (25) & -4.14 (16) & -0.302 (33) & --- \\
 \hline
$S_{2p}(Eq.\ref{eq:s2p})$ &0.306&0.0449&0.303&0.320&0.102& 		-0.478 (8) & 0.105 (7) &---   & -8.36 (22)	&---&--- \\
$S_{2p}(Eq.\ref{eq:s2p02})$ &0.265&0.0439&0.261&0.390&0.289& 		-0.360 (19) & -0.00125 (1794) & 0.00736 (111)      & -8.55 (21)	& --- & --- \\
$S_{2p}(Eq.\ref{eq:s2p03})$ &0.264&0.0355&0.261&0.267&0.0527& 	-0.361 (5) & --- & 0.00729 (34) & -8.55 (20)	& --- & --- \\
\bottomrule[0.65pt]
\end{tabular*}

\begin{tablenotes}
	\footnotesize
	\item[*] Calculated uncertainties for these nuclides with experimentally known data.
	\item[\#] Calculated uncertainties for these nuclides without experimentally known data.
%	\item[m] The middle region.
%	\item[o] The outer region.
%	\item[a2m] The middle region estimated by parameters obtained from all region.
%	\item[a2o] The outer region estimated by parameters obtained from all region.
%	\item[m2o] The outer region estimated by parameters obtained from the middle region.
%	\item[o2m] The middle region estimated by parameters obtained from the outer region.
\end{tablenotes}
\end{threeparttable}
\end{center}
\end{table*}

\begin{table*}
\caption{\label{tab:pdtBE}Comparison of predicted ground-state binding energies with those in AME2016 \cite{wang2017ame2016} for the even-$Z$ nuclides with $N\geqslant Z-6$ and odd-$Z$ nuclides with $N\geqslant Z-5$. The unit of energy is in MeV.}
\begin{center}
\footnotesize \tabcolsep 2pt 
\begin{tabular*}{\textwidth}{cccccccc|cccccccc}
\toprule
Nucl.     & Eq. (\ref{eq:yuan}) & Eq. (\ref{eq:chen}) & AME2016    & Nucl.      & Eq. (\ref{eq:yuan}) & Eq. (\ref{eq:chen}) & AME2016  & Nucl.     & Eq. (\ref{eq:yuan}) & Eq. (\ref{eq:chen}) & AME2016    & Nucl.      & Eq. (\ref{eq:yuan}) & Eq. (\ref{eq:chen}) & AME2016        \\
\hline
$^{62}$Ge & 516.994                              & 517.672                              & 517.142  & $^{63}$As & 515.644                              & 516.200                              & 516.159      & $^{84}$Tc & 681.878                              & 681.420                              & 682.080  & $^{85}$Tc & 698.130                              & 697.735                              & 698.275                    \\
$^{64}$As & 529.876                              & 530.363                              & 530.304  & $^{64}$Se & 516.288                              & 516.717                              & 516.672    & $^{86}$Tc & 712.049                              & 711.722                              & 712.080  & $^{82}$Ru & 631.847                              & 631.119                              &                            \\
$^{65}$Se & 530.673                              & 531.043                              & 531.050  & $^{66}$Se & 547.138                              & 547.453                              & 547.800     & $^{83}$Ru & 647.709                              & 647.043                              &               & $^{84}$Ru & 666.221                              & 665.622                              &                               \\
$^{65}$Br & 513.356                              & 513.653                              &                & $^{66}$Br & 528.900                              & 529.148                              &                    & $^{85}$Ru & 681.555                              & 681.025                              & 682.550 & $^{86}$Ru & 698.695                              & 698.238                              & 699.438               \\
$^{67}$Br & 545.496                              & 545.698                              & 546.184  & $^{68}$Br & 559.822                              & 559.981                              & 560.252      & $^{87}$Ru & 712.819                              & 712.439                              & 713.313 & $^{88}$Ru & 729.081                              & 728.781                              & 730.224                   \\
$^{66}$Kr & 512.830                              & 512.991                              &                & $^{67}$Kr & 528.600                              & 528.720                              &                  & $^{89}$Ru & 741.323                              & 741.106                              & 742.171 & $^{85}$Rh & 662.535                              & 661.849                              &                     \\
$^{68}$Kr & 546.273                              & 546.357                              &                & $^{69}$Kr & 560.616                              & 560.667                              & 561.177    & $^{86}$Rh & 678.669                              & 678.062                              &               & $^{87}$Rh & 696.063                              & 695.538                              &                             \\
$^{70}$Kr & 577.440                              & 577.462                              & 577.920  & $^{69}$Rb & 543.093                              & 543.054                              &                  & $^{88}$Rh & 711.244                              & 710.805                              & 711.920 & $^{89}$Rh & 727.788                              & 727.438                              & 728.999                  \\
$^{70}$Rb & 558.518                              & 558.456                              &               & $^{71}$Rb & 575.800                              & 575.718                              & 576.165    & $^{90}$Rh & 742.871                              & 742.614                              & 742.950 & $^{91}$Rh & 757.236                              & 757.076                              & 757.848                   \\
$^{72}$Rb & 590.462                              & 590.363                              & 590.544 & $^{73}$Rb & 606.793                              & 606.681                              & 606.338     & $^{86}$Pd & 661.508                              & 660.732                              &               & $^{87}$Pd & 677.532                              & 676.843                              &                              \\
$^{70}$Sr & 542.659                              & 542.493                              &                & $^{71}$Sr & 558.180                              & 558.000                              &                     & $^{88}$Pd & 696.041                              & 695.444                              &               & $^{89}$Pd & 711.176                              & 710.674                              &                              \\
$^{72}$Sr & 576.369                              & 576.178                              &                & $^{73}$Sr & 590.968                              & 590.770                              & 591.446      & $^{90}$Pd & 729.042                              & 728.639                              & 730.170 & $^{91}$Pd & 743.462                              & 743.162                              & 744.471                 \\
$^{74}$Sr & 608.019                              & 607.817                              & 608.354  & $^{73}$Y  & 573.321                              & 573.017                              &                    & $^{92}$Pd & 760.500                              & 760.305                              & 761.116 & $^{93}$Pd & 773.310                              & 773.224                              & 773.667                 \\
$^{74}$Y  & 588.947                              & 588.644                              &                & $^{75}$Y  & 606.148                              & 605.851                              & 606.675     & $^{89}$Ag & 692.519                              & 691.845                              &              & $^{90}$Ag & 708.699                              & 708.130                              &                              \\
$^{76}$Y  & 621.059                              & 620.771                              & 621.376  & $^{77}$Y  & 636.978                              & 636.702                              & 637.406      & $^{91}$Ag & 726.633                              & 726.172                              &              & $^{92}$Ag & 742.292                              & 741.943                              & 742.900                     \\
$^{78}$Y  & 650.758                              & 650.498                              & 651.222  & $^{74}$Zr & 572.619                              & 572.197                              &                    & $^{93}$Ag & 759.299                              & 759.064                              & 760.089 & $^{94}$Ag & 774.707                              & 774.591                              & 774.372                \\
$^{75}$Zr & 588.291                              & 587.879                              &                & $^{76}$Zr & 606.415                              & 606.018                              &                    & $^{95}$Ag & 789.642                              & 789.648                              & 789.640 & $^{90}$Cd & 691.561                              & 690.805                              &                         \\
$^{77}$Zr & 621.419                              & 621.041                              & 622.237  & $^{78}$Zr & 638.355                              & 637.998                              & 639.132       & $^{91}$Cd & 707.865                              & 707.223                              &               & $^{92}$Cd & 726.644                              & 726.120                              &                             \\
$^{79}$Zr & 652.222                              & 651.891                              & 653.093  & $^{80}$Zr & 668.167                              & 667.864                              & 668.800       & $^{93}$Cd & 742.306                              & 741.903                              &               & $^{94}$Cd & 760.402                              & 760.123                              & 761.306                \\
$^{77}$Nb & 603.349                              & 602.848                              &               & $^{78}$Nb & 619.281                              & 618.807                              &                    & $^{95}$Cd & 775.394                              & 775.242                              & 775.865 & $^{96}$Cd & 792.748                              & 792.728                              & 792.864             \\
$^{79}$Nb & 636.529                              & 636.086                              & 637.214 & $^{80}$Nb & 651.777                              & 651.369                              & 652.080      & $^{97}$Cd & 805.940                              & 806.055                              & 805.779 & $^{93}$In & 723.213                              & 722.621                              &                        \\
$^{81}$Nb & 667.637                              & 667.267                              & 668.088 & $^{82}$Nb & 681.565                              & 681.237                              & 681.830     & $^{94}$In & 739.896                              & 739.434                              &                & $^{95}$In & 757.960                              & 757.632                              &                    \\
$^{78}$Mo & 602.231                              & 601.620                              &              & $^{79}$Mo & 618.297                              & 617.723                              &                   & $^{96}$In & 774.172                              & 773.981                              & 774.432  & $^{97}$In & 791.454                              & 791.403                              & 791.811               \\
$^{80}$Mo & 636.449                              & 635.916                              &              & $^{81}$Mo & 651.571                              & 651.082                              & 652.698     & $^{98}$In & 806.975                              & 807.069                              & 806.540  & $^{99}$In & 822.625                              & 822.866                              & 822.096                \\
$^{82}$Mo & 668.686                              & 668.245                             & 669.366 & $^{83}$Mo & 683.032                              & 682.642                              & 683.422      & $^{94}$Sn & 722.013                              & 721.348                              &               & $^{95}$Sn & 738.737                              & 738.211                              &                          \\
$^{84}$Mo & 698.914                              & 698.578                             & 699.300 & $^{81}$Tc & 633.033                              & 632.405                              &                     & $^{96}$Sn & 757.639                              & 757.257                              &               & $^{97}$Sn & 773.837                              & 773.601                              &                             \\
$^{82}$Tc & 649.053                              & 648.478                              &                & $^{83}$Tc & 666.556                              & 666.038                              & 667.569      & $^{98}$Sn & 792.195                              & 792.109                              &               & $^{99}$Sn & 807.901                              & 807.968                              & 807.840   \\
\bottomrule
\end{tabular*}
\end{center}
\end{table*}

\subsection{Binding Energy}

	The binding energy formulas Eqs. (\ref{eq:1})-(\ref{eq:E.Caurier}) are applied to the 221 nuclides with measured binding energies via the Bootstrap framework. The obtained parameters and uncertainties are summarized in TABLE \ref{tab:stat}. As mentioned in Sec. \ref{sec:models}, the effect of Coulomb interaction is included in terms $a$ and $b$. The repulsion between protons will decrease the single particle energy of proton orbit and the energy released when they form a nucleus, which is consistent with the negative values of $a$ and $b$. In Eq. (\ref{eq:yuan}), the residual is summarized to the competition between the neutron shell effect and the isospin effect, which are represented through the square of valence neutrons and that of difference between valence protons and neutrons. As parameters $c$ and $d$ of Eq. (\ref{eq:yuan}) are with similar scale but opposite sign, these two effects compensate for each other when it is far away from the proton dripline, this also leads to the equivalence between Eqs. (\ref{eq:yuan}) and (\ref{eq:chen}). The large systematic and statistical uncertainties of Eq. (\ref{eq:1}), which are about twice larger than those of the rest, demonstrate the deficiency of Eq. (\ref{eq:1}). Eq. (\ref{eq:E.Caurier}) achieves the best performance regarding the total and systematic uncertainties, but at the price of 6 parameters, which results in great extrapolating weakness. To control the extrapolating uncertainty in the prediction, it is recommended to consider less parameters.

\begin{figure}
\includegraphics[width=.4\textwidth]{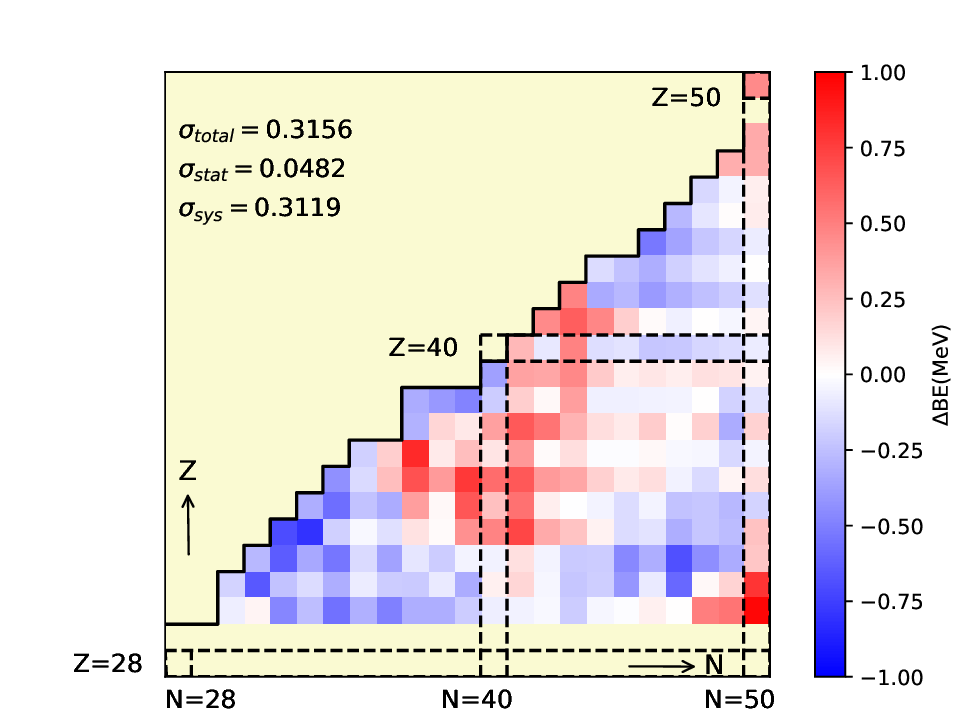}
\includegraphics[width=.4\textwidth]{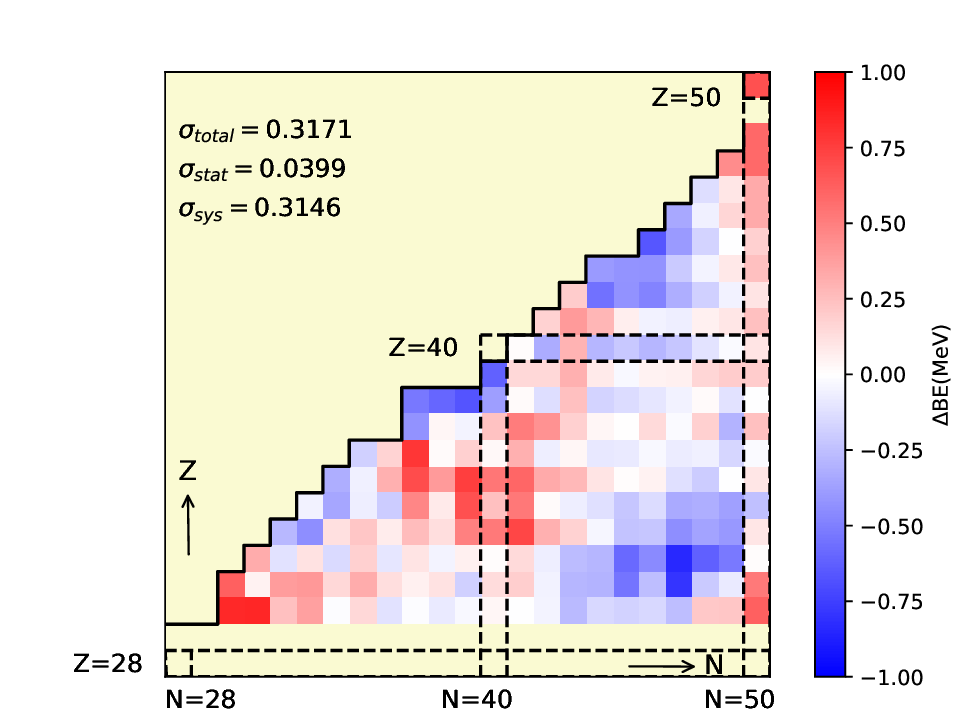}
\caption{\label{fig:rsd_BE}Distribution of the mean residual of each nuclide for binding energy formulas Eqs. (\ref{eq:yuan}) and (\ref{eq:chen}) in order. The dark solid line shows the measurement boundary of binding energy in AME2016 \cite{wang2017ame2016}.}
\end{figure}

\begin{figure}
\includegraphics[width=.4\textwidth]{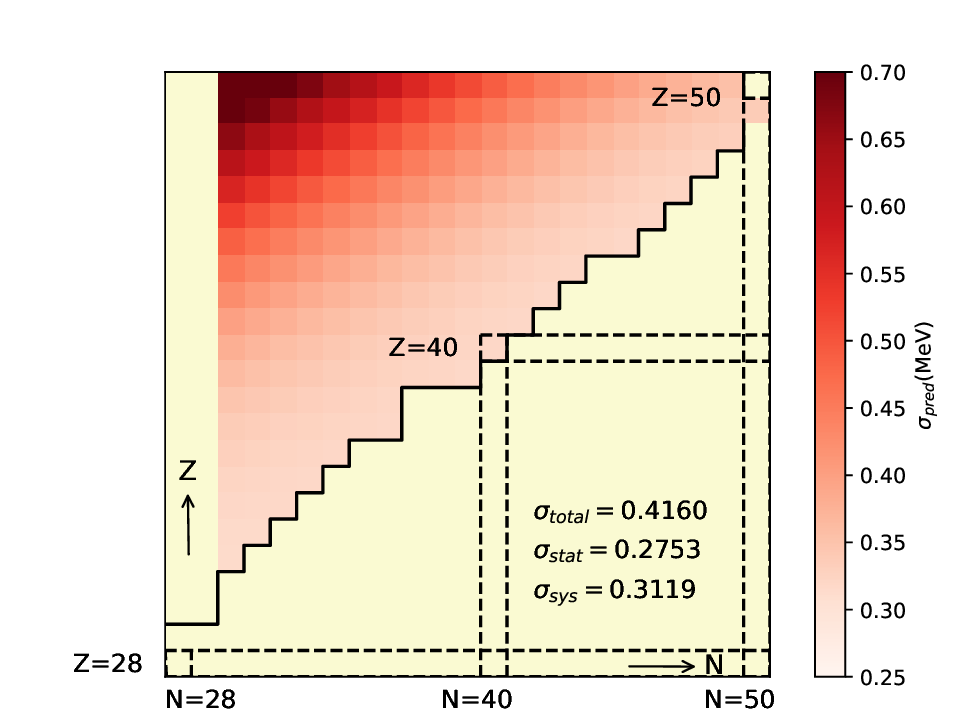}
\includegraphics[width=.4\textwidth]{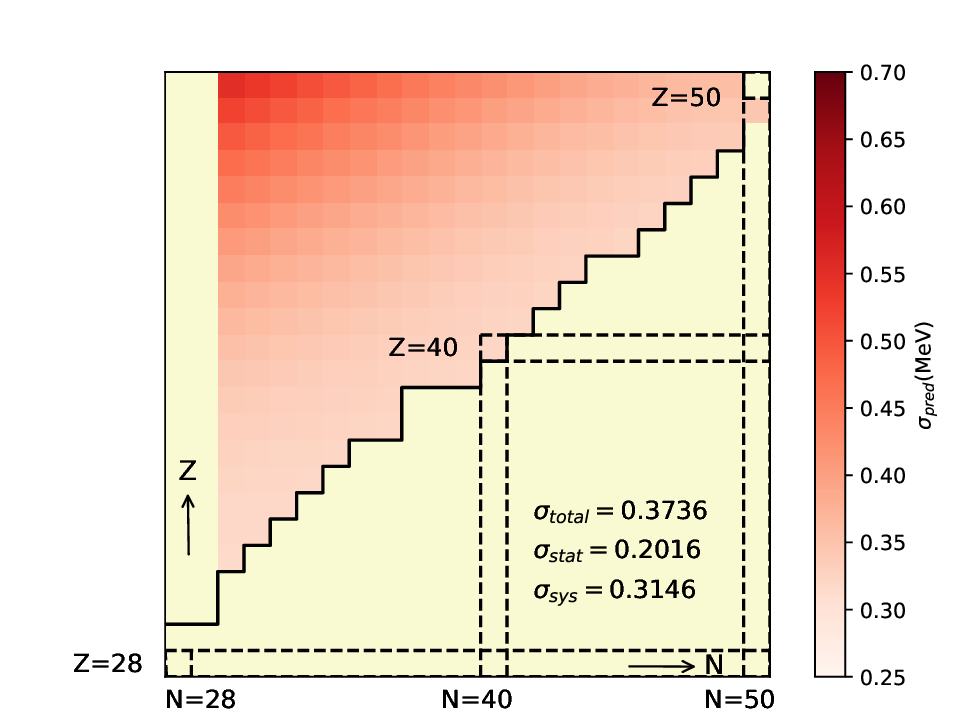}
\caption{\label{fig:uncer_BE}Distribution of predictive uncertainty of each nuclide for binding energy formulas Eqs. (\ref{eq:yuan}) and (\ref{eq:chen})  in order. The dark solid line shows the measurement boundary of binding energy in AME2016 \cite{wang2017ame2016}.}
\end{figure}

	FIG. \ref{fig:rsd_BE} shows the average of residuals for each nuclide with experimentally determined binding energy evaluated by AME2016 in the region of $30 \leqslant Z\text{, } N \leqslant50$. The values of residuals are presented by the gradation of color. Starting from the white, the redder means more positive and bluer means more negative. No data have residuals larger than $3\hat{\sigma}_{total}$ or smaller than $-3\hat{\sigma}_{total}$. The extrapolation power is delineated in FIG. \ref{fig:uncer_BE} by the recomposed uncertainty deduced from Eq. (\ref{eq:sigma_pred_all}). Both formulas exhibit uncertainties which are small near the reached binding energy boundary and increase when moving away. Our predictions of binding energies are mostly smaller than the extrapolation of AME2016 shown in TABLE \ref{tab:pdtBE}. This is consistent with the fact that nuclei near the dripline are less bound than nuclei near the stability line, since the extrapolation of AME2016 were obtained under an assumption of smooth mass surface \cite{huang2017ame2016}. Besides, more nuclides bound under energy criterion are predicted in the present work, which is waiting for experimental examination.

	We also try another interaction, jj44bpn \cite{brown2014shell}, in the same model space, but it does not perform well on the description of binding energies. For quick comparison, $E_{BE, SM}$ of nuclides with $45\leqslant N \leqslant 50$ or $30\leqslant Z\leqslant33$ are calculated through these two interactions. When applying to Eqs.(\ref{eq:1})-(\ref{eq:E.Caurier}), jj44bpn still has larger rms of residuals compared with JUN45 as listed in TABLE \ref{tab:comp-jun45-jj44bpn}. Since the jj44bpn does not show better description of binding energy than JUN45, the corresponding results from jj44bpn are not further discussed in the present study.

\begin{table}
\begin{center} 
\footnotesize
\caption{\label{tab:comp-jun45-jj44bpn}The rms of residuals when JUN45 and jj44bpn are applied to Eqs. (\ref{eq:1})-(\ref{eq:E.Caurier}) among nuclides with $N \geqslant 45$ or $Z \leqslant 33$.}
\begin{tabular}{ccccc}
\toprule
		& Eq. (\ref{eq:1})& Eq. (\ref{eq:yuan}) & Eq. (\ref{eq:chen}) & Eq. (\ref{eq:E.Caurier})\\
\hline
JUN45 		& 0.586		& 0.249			&0.292 &0.249 \\
jj44bpn		& 0.722		& 0.873			&0.848 &0.700 \\
\bottomrule
\end{tabular}
\end{center}
\end{table}%

	When constructing JUN45 interaction, nuclides with $N < 46$ and $Z > 33$ are excluded due to their deformations, that the model space may not be sufficient to describe \cite{honma2009new}. In order to investigate and evaluate the description of nuclides in the middle region, the dataset is divided into two parts: 1) $N < 45 $ \& $ Z > 33$; 2) $N \geqslant 45$ or $Z \leqslant 33$. The Bootstrap framework is performed separately on these two subsets. The quantification of the deformation effect comes from a cross-extrapolation estimation. As listed in TABLE \ref{tab:cross-estimation}, comparing with the self-estimated result, the systematic uncertainty of the middle region increases largely when calculated by parameters estimated with the outer region, and vice versa. This does show the difference between nuclides in these two regions. However, the proposed corrections Eqs.(\ref{eq:yuan}) and (\ref{eq:chen}) lead a trade-off when the full dataset is considered. Uncertainties of the middle region, estimated by parameters fitted to the whole dataset, deviate less than 0.08 MeV comparing with the self-estimated result. Thus it is reasonable to describe the nuclides in the middle region through the present framework.

\begin{table}
\begin{center}
\footnotesize 
\caption{\label{tab:cross-estimation}Comparison of the decomposed uncertainties estimated by parameters obtained from the middle region ($N < 45 $ \& $ Z > 33$), the outer region ($N \geqslant 45$ or $Z \leqslant 33$) and the whole region. Parameters of Eq. (\ref{eq:yuan}) obtained from the middle region are $a=-9.62 (3)$, $b=-0.0858 (37)$, $c=0.0175 (18)$, $d=-0.0108 (44)$; that from the outer region are $a=-9.53 (2)$, $b=-0.0968 (8)$, $c=0.0229 (5)$, $d=-0.0250 (9)$. Parameters of Eq. (\ref{eq:chen}) obtained from the middle region are $a=-9.19 (2)$, $b=-0.0873 (21)$, $c=-0.0276 (13)$; that from the outer region are $a=-9.10 (1)$, $b=-0.0932 (7)$, $c=-0.0305 (7)$. Parameters obtained from the whole region are taken from TABLE \ref{tab:stat}.}
\begin{tabular}{cccccc}
\toprule
&objective region & parameter region & $\sigma_{total}$\tnote{*} & $\sigma_{stat}$\tnote{*} & $\sigma_{sys}$\tnote{*} \\
\hline
\multirow{6}{*}{Eq. (\ref{eq:yuan})} &\multirow{3}{*}{middle}	& middle &0.317&0.0849&0.305\\
	&					& outer	&0.557&0.0569&0.554  \\
	&					& whole	&0.379&0.0437&0.377\\
	&\multirow{3}{*}{outer}	& outer	&0.254&0.0473&0.250\\
	&				&middle	&0.867&0.305&0.812 \\
	&					&whole	&0.286&0.0499&0.281    \\
 \hline
\multirow{6}{*}{Eq. (\ref{eq:chen})}&\multirow{3}{*}{middle}	& middle  	&0.327&0.0768&0.317 \\
	&					& outer 	&0.382&0.0724&0.375 \\
	&					& whole 	&0.355&0.0431&0.352 \\
	&\multirow{3}{*}{outer}	& outer 	&0.295&0.0426&0.292 \\
	&				&middle	&0.461&0.153 &0.434 \\
	&					&whole 	&0.300&0.0385&0.298\\
\bottomrule
\end{tabular}
\end{center}
\end{table}

	Random perturbations in gaussian form with $\sigma$ of 20\% ($\mathcal{N}(x, \sigma=0.2x)$) are applied to the 133 two-body matrix elements (TBMEs). If the uncertainty of the JUN45 interaction remains large, a random perturbation would have a significant probability of obtaining better results in binding energies. The region is narrowed to the 128 nuclides whose total valence particles and holes are less than 13, corresponding to $A\leqslant 68$ or $A\geqslant 88$ or $N\geqslant Z+10$. Specifically, the application of perturbation to the JUN45 interaction is divided into three groups according to the TBMEs being changed: only the diagonal TBMEs (D), only the non-diagonal TBMEs (ND), and both of them (D+ND). Then, the energies relative to the core resulting from the perturbed interactions are put into the binding energy formulas to make a fit and the rms of residuals is calculated. For each group, the perturbation is applied 50 times.
	
\begin{figure}
\includegraphics[width=.237\textwidth]{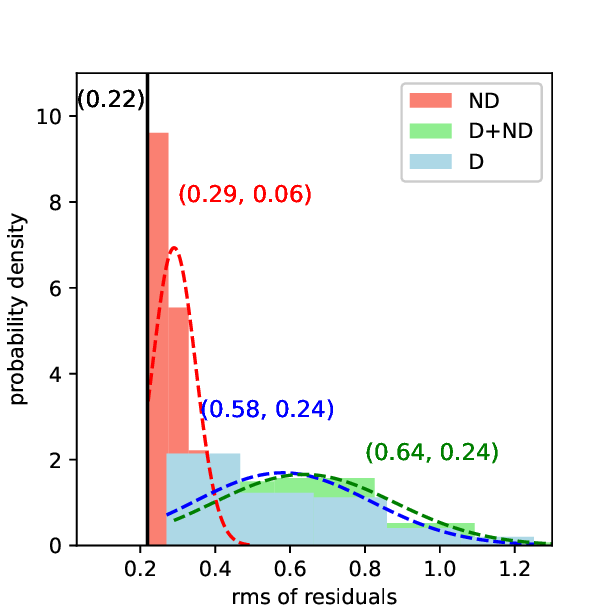}
\includegraphics[width=.237\textwidth]{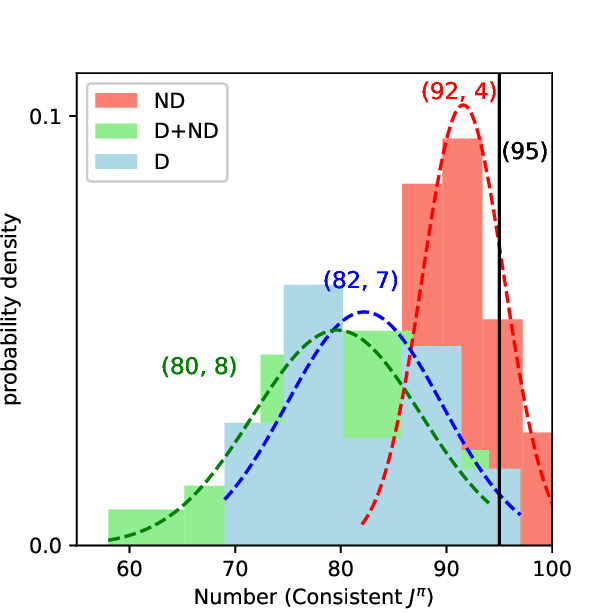}
\caption{\label{fig:perturbation}Distribution of the rms of residuals and the number of consistent $J^\pi$ for perturbations applied to the JUN45 interaction. The dash lines are the fitted normal distribution, of which the parameters, mean and standard deviation, are listed in the nearby parentheses. The vertical black solid line denotes the value when no perturbation is applied, and the black number in parentheses is its corresponded value.}
\end{figure}

	As listed in TABLE \ref{tab:comp-jun45-ptbt} and shown in FIG. \ref{fig:perturbation}, the average of rms of residuals is larger than that without perturbation for each binding energy formula. But the influence of perturbation applied to the non-diagonal TBMEs is weak, since its average of rms of residuals exceeds the unperturbed one just around 0.07 MeV and its distribution is narrow, as drawn in FIG. \ref{fig:perturbation}.
	
\begin{table}
\begin{center}
\footnotesize
\caption{\label{tab:comp-jun45-ptbt}Comparison of the rms of residuals when no perturbation is applied to the JUN45 interaction, with the average and standard deviation of rms of residuals when perturbation is applied to D, ND and D+ND of the JUN45 interaction among nuclides whose number of valence particles and holes is less than 13.}
\begin{tabular}{ccccc}
\toprule
		& Eq. (\ref{eq:1})& Eq. (\ref{eq:yuan}) & Eq. (\ref{eq:chen}) & Eq. (\ref{eq:E.Caurier})\\
\hline
JUN45 	& 0.651		& 0.219			&0.292 &0.216 \\
D		& 0.778 (201)	& 0.580 (236)		&1.119 (421) &0.464 (159)\\
ND		& 0.680 (57)	& 0.290 (58)		&0.359 (62) &0.259 (36) \\
D+ND	& 0.787 (181)	& 0.638 (241)		&0.979 (388) &0.505 (148)\\
\bottomrule
\end{tabular}
\end{center}
\end{table}%
	
	The rms is sensitive to the diagonal TBMEs, as shown through the wide expansion of distribution in case D and D+ND. Moreover, the rms of residuals without perturbation drops out of 1.5 (1.75) $\sigma$ of the distribution when the perturbation is applied to the diagonal (all) TBMEs, which implies the significant influence of perturbation and the well fitted diagonal TBMEs of JUN45 interaction. In addition, the number of nuclides, whose spin and parity ($J^\pi$) resulting from the perturbed interaction are consistent with that from the observation, is counted. The distribution of the consistency is drawn in the right panel of FIG. \ref{fig:perturbation}.  It shows similar results to the perturbation cases as for rms. Therefore, the uncertainty of JUN45 interaction in energy calculation is shown to be insignificant without perturbation.
	
	The spin and parity of ground state calculated by the JUN45 interaction ($J^\pi_{\text{JUN45}}$) are also compared with those ($J^\pi_{exp}$) from the NNDC. Among the 198 nuclides with determined $J^\pi_{exp}$ , there are 137 nuclides whose  $J^\pi_{\text{JUN45}}$ is consistent with $J^\pi_{exp}$. Besides, there are 54 nuclides with undetermined $J^\pi_{exp}$. This rarely influences the description of the absolute value of binding energy, which is a bulk property of nucleus. Generally, nuclear mass models, e.g., the developed semi-empirical droplet model \cite{myers1966nuclear,myers1976development}, the finite-range droplet model \cite{moller1981atomic, moller1981nuclear, moller1988nuclear, moller1988nuclear2, moller1993nuclear, moller2016nuclear} and the application of Hartree-Fock-Bogoliubov \cite{goriely2013hartree, wang2014surface} do not concentrate on the spin and parity.
	
\subsection{Separation Energies, Stable Possibility and Partial Half-lives}

	As Eqs. (\ref{eq:yuan}) and (\ref{eq:chen}) have good agreement with the experimental binding energies, the deduced formulas are accordingly investigated for the separation energies of the last proton and the last two protons, which are expressed by Eqs. (\ref{eq:sp})-(\ref{eq:s2p03}). The Bootstrap framework is applied directly to the separation energy formulas rather than calculating separation energies by binding energy formulas, which can avoid error propagation. Respectively, Eqs. (\ref{eq:sp}), (\ref{eq:sp02}), (\ref{eq:sp03}) and Eqs. (\ref{eq:s2p}), (\ref{eq:s2p02}), (\ref{eq:s2p03}) are applied on the 198 $S_{p}$ and 178 $S_{2p}$ evaluated in AME2016.

\begin{figure}
\includegraphics[width=.4\textwidth]{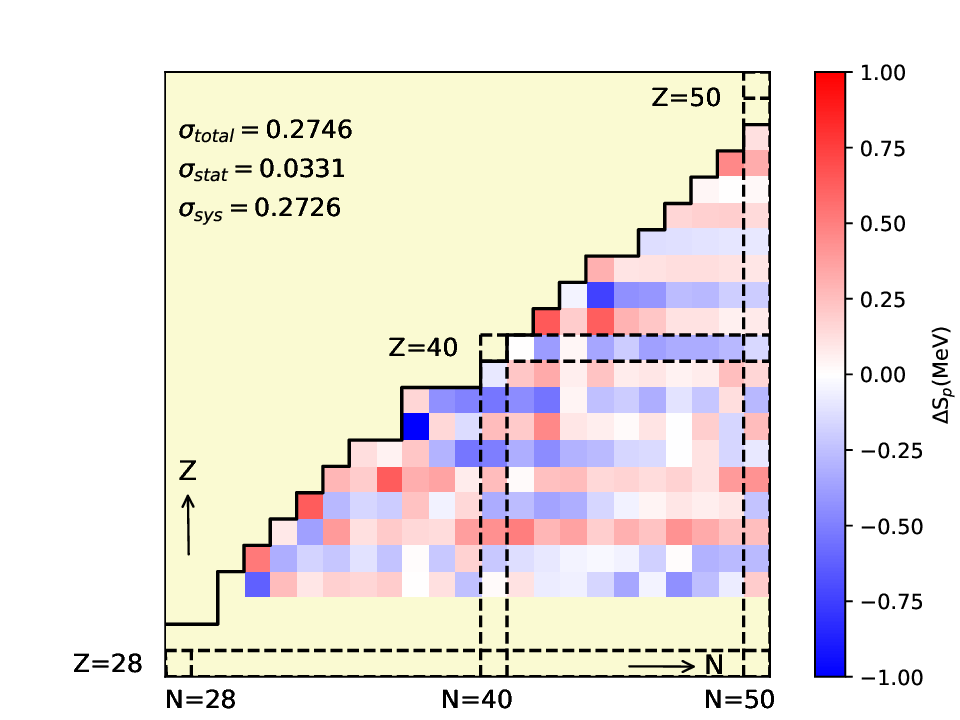}
\includegraphics[width=.4\textwidth]{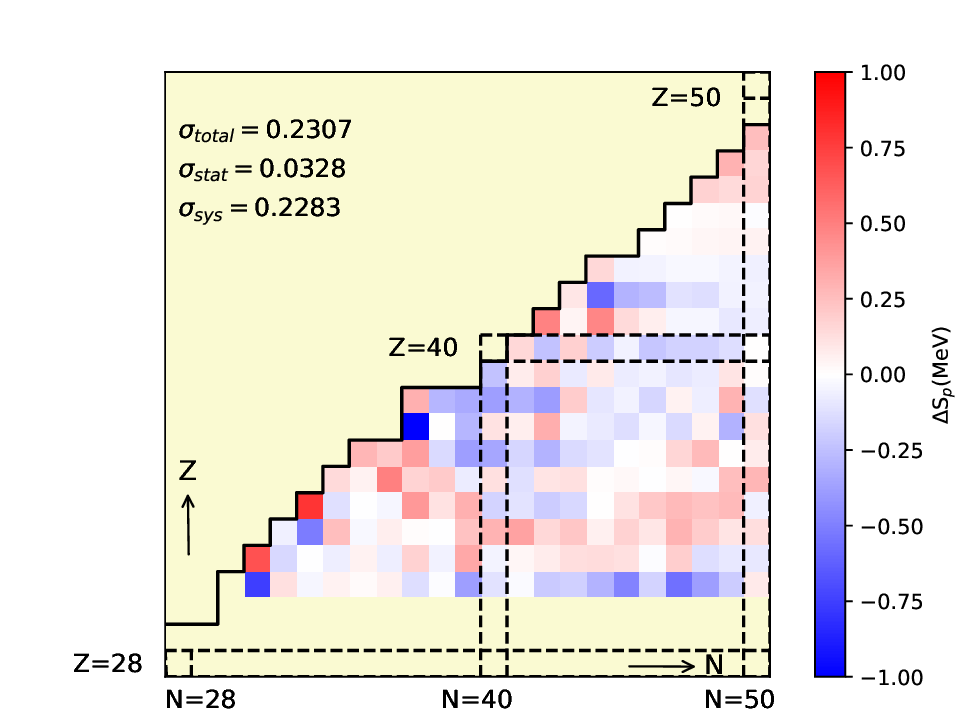}
\caption{\label{fig:rsd_S}Distribution of the mean residual of each nuclide for $S_{p}$ formulas Eqs. (\ref{eq:sp03}) and (\ref{eq:sp04}) in order.}
\end{figure}

	The fitting values of parameters $a$, $c$, and $d$ of the $S_{2p}$ formulas are almost a factor of two larger than those of the corresponding parameters for the $S_{p}$ formulas. This approximated double relation is mainly caused by the nature of subtracting respectively two and one protons in calculating $S_{2p}$ and $S_{p}$. Eq. (\ref{eq:sp03}) tends to overestimate odd-$Z$ nuclides but underestimate even-$Z$ nuclides, which indicates the paring strength may not be well described in the shell-model calculations. It enlightens to introduce an estimation for such extra energy as
\begin{equation}
\begin{split}
	S_{p}(Z, N)=E_{BE, SM}(Z, N) - E_{BE, SM}(Z-1, N) \\
	 + aZ + c(Z-N)^{2} + d + e\delta_{Z},
	\label{eq:sp04}
\end{split}
\end{equation}
where $\delta_{Z}=0$ if $Z$ is even and $\delta_{Z}=1$ if $Z$ is odd. The correction indeed reduces the uncertainty as listed in TABLE \ref{tab:stat} and, as shown in FIG. \ref{fig:rsd_S},  smoothens the residual distribution.

\begin{figure}
\includegraphics[width=.4\textwidth]{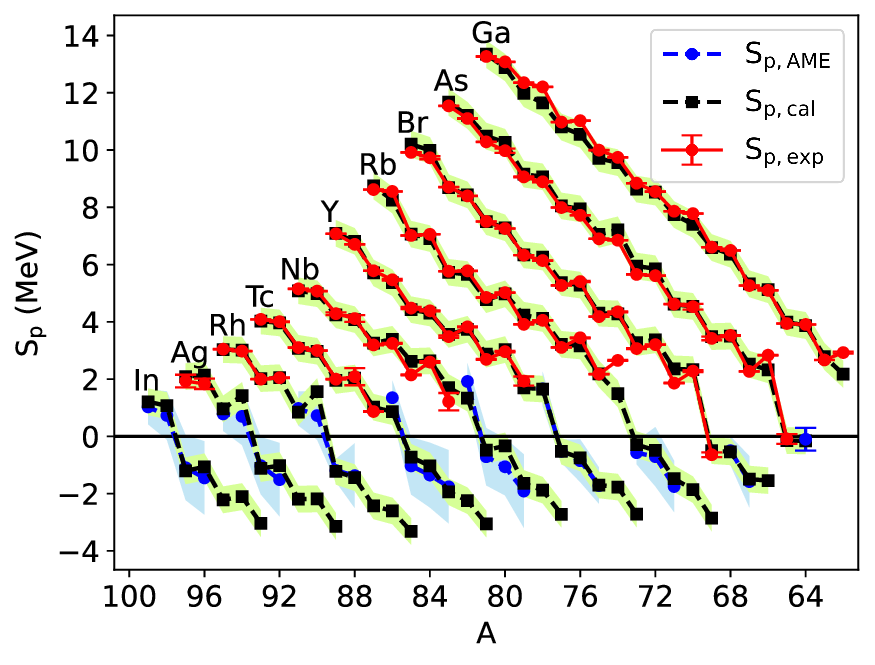}
\includegraphics[width=.4\textwidth]{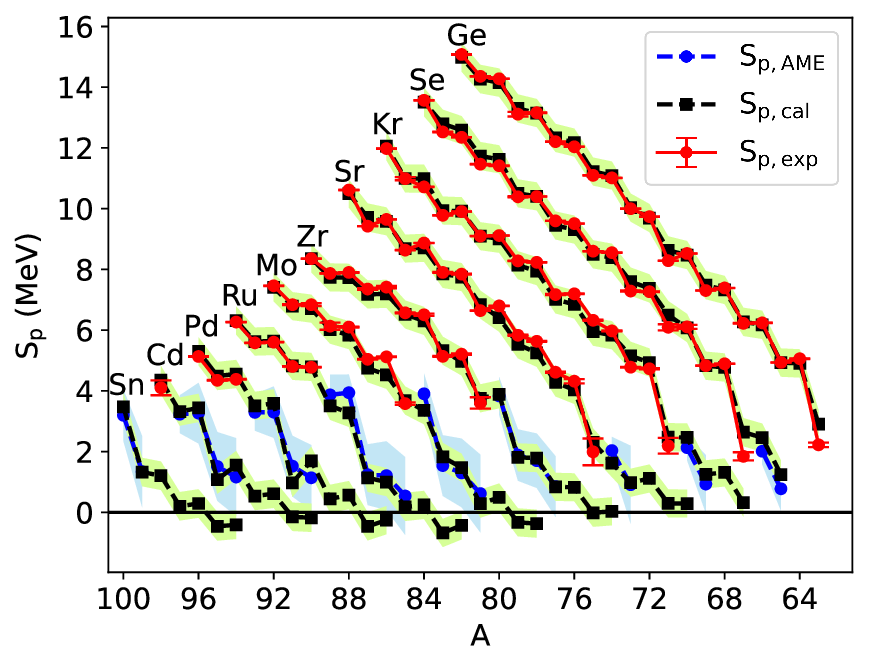}
\caption{\label{fig:compare_Sp_1} $S_{p}$ calculated in this work and that determined and extrapolated in AME2016. $S_{p, cal}$ is the values calculated through Eq. (\ref{eq:sp04}), $S_{p, exp}$ is the values determined experimentally in AME2016, and $S_{p, AME}$ is the extrapolated values in AME2016. The green region, the red error bars and the blue region denote the 2$\sigma$ uncertainties for $S_{p,cal}$, $S_{p,exp}$ and $S_{p,AME}$, respectively.}
\end{figure}

\begin{figure}
\includegraphics[width=.4\textwidth]{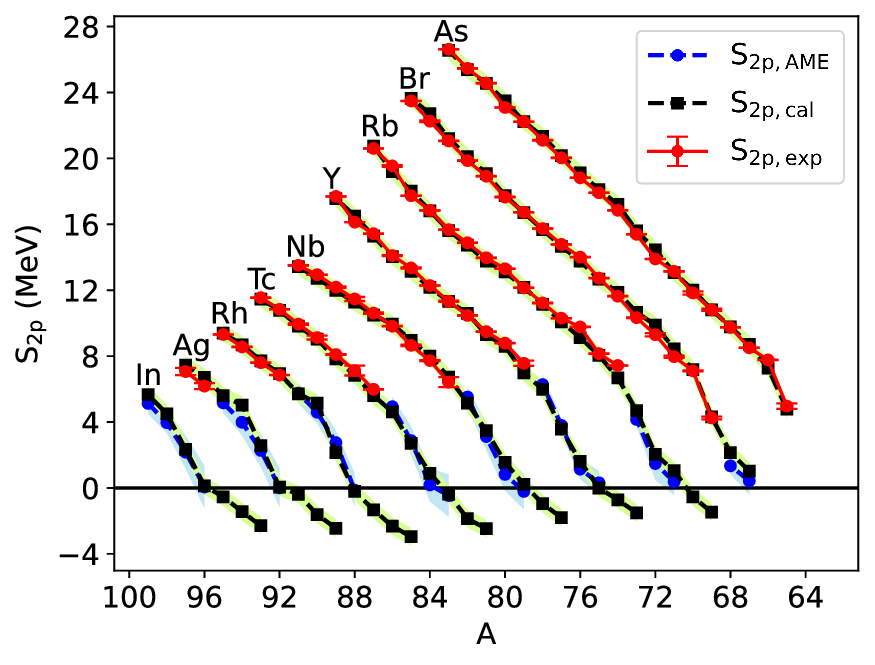}
\includegraphics[width=.4\textwidth]{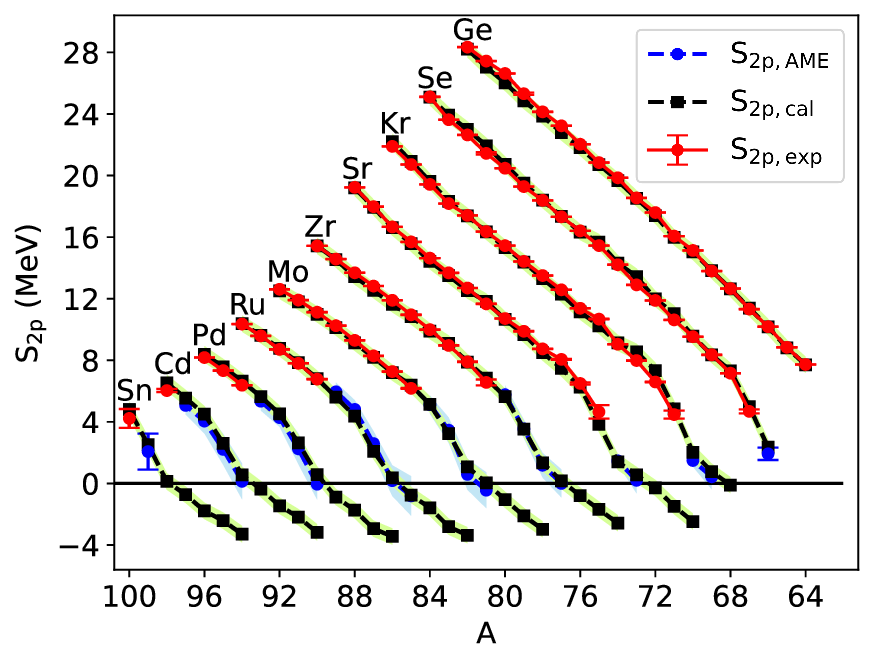}
\caption{\label{fig:compare_S2p_1} $S_{2p}$ calculated in this work and that determined and extrapolated in AME2016. See similar caption of FIG. \ref{fig:compare_Sp_1}.}
\end{figure}

	Beyond the proton(s) boundary discovered experimentally in AME2016, the separation energies are also calculated theoretically. FIGs. \ref{fig:compare_Sp_1} and \ref{fig:compare_S2p_1} compare respectively the values of $S_{p}$ and $S_{2p}$ calculated through Eqs. (\ref{eq:sp04}) and (\ref{eq:s2p03}) and those of AME2016. The calculated values agree well with the experimental data, which are mostly located within the 2$\sigma$ range of calculations. The present calculations generally agree with the extrapolated values in AME2016, but with smaller uncertainties. This provides important inputs to the simulation of nuclear astrophysics and have significant impact on the understanding of $p$-process nuclei and their solar abundance, since the $S_p$, taken by Pruet $et \ al.$ for the $\nu p$-process reaction flow going through the Zn-Sn regions to explore the production of the $p$ nuclei in the neutrino-driven wind from a young neutron star in Type II supernovae \cite{pruet2006nucleosynthesis}, are with very large extrapolated uncertainty. They found synthesis of $p$-rich nuclei could be reached up to $^{102}$Pd, although their calculations do not show efficient production of $^{92}$Mo. If the entropy of these ejecta were increased by a factor of 2, the synthesis could extend to $^{120}$Te. Still larger increases in entropy, which might reflect the role of magnetic fields or vibrational energy input neglected in the hydrodynamical model, resulted in the production of nuclei up to $A\approx170$.
	
	For isotope chains of Ge, As, Se, Br, Kr and Sr, the observed $S_{p}$ decreases sharply when $Z$ exceeds $N$. Both extrapolations in the present work and in AME2016 present such characteristics for other isotopes. As to odd-Z nuclides (in the top of FIG. \ref{fig:compare_Sp_1}), such gap leads to the single proton dripline while the pairing of protons makes the single proton dripline of even-Z nuclides farther to reach. This phenomenon exists also for $S_{2p}$ as shown in FIG. \ref{fig:compare_S2p_1}. Sharp decrease is shown for As, Se, Br, Kr, Sr and predicted for other isotopes when $Z$ exceeds $N$ but is smoothed because of the correlation between the two emitted protons. Thus, the proton dripline tends to be extended beyond $Z=N$ line by the competition between paring of protons and $Z/N$ symmetry.

	The resampling process in the Bootstrap framework estimates the parameter space of Eqs. (\ref{eq:sp04}) and (\ref{eq:s2p03}). Simultaneously, it also estimates the distribution of $S_p$ and $S_{2p}$ for each nuclide. Hence, the possibilities of $S_{p}<0$ and $S_{2p}<0$ for each nuclide are obtained by integrating the estimated normalized distribution. As listed in TABLE \ref{tab:pdt}, the \emph{p}- and 2\emph{p}- driplines are predicted based on the present calculations.

	30 candidates, being not stable but bound against both $p$-emission and 2$p$-emission, are predicted in TABLE \ref{tab:pdt} under condition $P(S_{p}<0)<1\% \cap P(S_{2p}<0)<1\%$, which are marked in bold in TABLE \ref{tab:pdt}. Note that $^{100}_{50}$Sn is experimentally $S_{2p}$ known and bound. The present calculation suggests that the drip-lines pass over $Z=N$ line in this region, which may provide new ideas for waiting points in the path of nucleosynthesis. The separation energies and bound probabilities also provide opportunity to estimate the pure $p$-emitters and pure $2p$-emitters in the region $Z,N\in [32, 50]$. Under conditions $P(S_{p}<0)>99\% \cap P(S_{2p}<0)<1\%$ and $P(S_{p}<0)<1\% \cap P(S_{2p}<0)>99\%$, 9 nuclides (marked with $^\#$) and 3 nuclides (marked with $^\dagger$) are predicted to be pure $p$-emitters and pure $2p$-emitters, respectively.
	
	With the assumption that the emission is dominated by channel from the ground state to the ground state and ignoring the deformation, the corresponding decay half-lives are calculated with a simple law proposed by Qi \emph{et al.} \cite{qi2012effects}:
	\begin{equation}
	\begin{split}
		\log_{10} T_{1/2} = a\chi'+b\rho'+dl(l+1)/\rho'+c,
		\label{eq:logT}
	\end{split}
	\end{equation}
	with $\chi'=Z_{p}Z_{d}\sqrt{X/Q_{p}}$, $\rho'=\sqrt{XZ_{p}Z_{d}(A_{p}^{1/3}+A_{d}^{1/3})}$ and $X = A_{p}A_{d}/(A_{p}+A_{d})$, subscript $p$ and $d$ denoting a proton and a daughter nucleus, respectively. The parameters are refit to be $a=0.4559, b=-0.4272, d=2.5706, c=-23.08$ based on the data of nuclides with $N < 75$ taken from Ref. \cite{qi2012effects}. Here, $^{130}$Eu and $^{112}$Cs are removed because of their large experimental uncertainty and undetermined angular momentum. The uncertainty of separation energy within the $2\sigma$ range is accounted for, which denotes a 95\% confidence interval. The experimentally determined $J^\pi_{g.s.}$ is taken in the partial half-life prediction. If the experimental data is not available, $J^\pi_{g.s.}$ of its mirror partner is used, which is a feasible choice. As far as our knowledge, there is only one observed mirror-symmetry-violated case on $J^\pi_{g.s.}$ \cite{hoff2020mirror}. If $J^\pi_{g.s.}$ values of both mirror partners are not known, the calculated values are used for a referenced estimation. 

	In recent years, several nuclides in this region have been identified by the experiments. Although some nuclides were not discovered, the limit of their partial half-life could be estimated. The experimentally estimated range of half-life of $^{68}$Br, $^{72,73}$Rb, $^{81}$Nb, $^{85}$Tc, $^{89}$Rh and $^{93}$Ag are summarized in TABLE \ref{tab:pdt}. These are consistent with the calculated values. $^{93}_{47}$Ag$_{46}$ and $^{89}_{45}$Rh$_{44}$ were measured to be one-proton emitters but against two protons emission \cite{vcelikovic2016new}, which are qualitatively consistent with the extrapolation results presented in TABLE \ref{tab:pdt}.
	
	In 2017, the proton emission from the ground-state of $^{72}$Rb was measured for the first time \cite{suzuki2017discovery}. Here, it is predicted to be proton unbound but two protons bound in this work. Recently, the proton decay of $^{72}$Rb is explained by the lower deformation and lower angular momentum barrier \cite{sinclair2019half}. It is consistent with the present calculation under 95\% confidence intervals. Its isotope, $^{73}$Rb, has not been directly measured yet. But the upper limit of its half-life was deduced \cite{suzuki2017discovery}. For $^{73}$Rb, the calculated partial half-life locates beyond the experimental upper limit so far, but the lower limit estimated theoretically is in consistency with it. This is because the presently calculated $S_{p}$ value is small but with large uncertainties, which induces large uncertainties on the predicted partial half-life.

	$^{76}$Y was identified in Ref. \cite{sinclair2019half}. The $\beta^{+}$ decay with a partial half-life of a few ms was measured to be the predominant mode of $^{76}$Y which suggests that this nuclide is possibly proton bound or acquires a sufficient small proton decay width \cite{sinclair2019half}. In contrast, $^{76}$Y is predicted to be proton unbound in the present work, as shown in TABLE \ref{tab:pdt}. A similar conclusion was drawn in another calculation performed by Kaneko \emph{et al.} \cite{kaneko2013variation}.
	
	As for $^{68}_{35}$Br, its existence was discovered recently and half-life was estimated to be 50 ns \cite{wimmer2019discovery}. This is consistent with the partial half-life estimated in this work. We suggest further investigation of one proton emitter of $^{68}$Br.

\begin{longtable*}{cccccccccc}
\footnotesize \tabcolsep 3.0pt \vspace{-3mm} \\
\caption{\label{tab:pdt}Predicted $S_p$, $S_{2p}$ and possibility of negative $S_p$ and $S_{2p}$ for experimentally undetermined nuclides, respectively. The partial half-lives in unit of second are calculated for nuclides with $P>70\%$ through Eq. (\ref{eq:logT}). Half-lives of several ground state nuclides determined experimentally are also listed. Nuclides, being not stable but bound against both $p$-emission and 2$p$-emission, are marked in bold. Those predicted to be pure $p$-emitters and pure 2$p$-emitters are marked respectively with $^\#$ and $^\dagger$.}\\
%\begin{center}

%\begin{tabular*}{\textwidth}
\toprule
Nucl.			& $l_{p}$	&	$S_p$ (MeV)	& $P (S_p < 0)$ 		& $\log_{10}T_{cal}$			& $\log_{10}T_{exp}$	& $l_{2p}$ &	$S_{2p}$ (MeV) & $P (S_{2p} < 0)$ & $\log_{10}T_{cal}$                \\
\hline
\endfirsthead
\toprule
Nucl.			& $l_{p}$	&	$S_{p}$ (MeV)	& $P (S_p < 0)$ 		& $\log_{10}T_{cal}$			& $\log_{10}T_{exp}$	& $l_{2p}$ &	$S_{2p}$ (MeV) & $P (S_{2p} < 0)$ & $\log_{10}T_{cal}$                \\
\hline
\endhead
\bottomrule
\endfoot
$^{64}$As	& 1		&-0.157	& 75\%	& $8.6_{-18.2}$		&					&		&		&			& \\
\textbf{$^{65}$Se} &	&1.244	& 0\%	& 					& 					&		&		&			& \\
\textbf{$^{66}$Se} &	&2.460	& 0\% 	&					& 					&		&2.352	& 0\% 		& \\
$^{66}$Br		& 1		&-1.545	& 100\% 	& $-15.8_{-1.5}^{+2.4}$ 	&                       			&		&		&			&                           \\
$^{67}$Br$^{\rm \#}$  & 1&-1.491	& 100\%	& $-15.6_{-1.6}^{+2.6}$ 	&                       			&		&1.019	& 0\%		&                           \\
$^{68}$Br$^{\rm \#}$  & 1&-0.540	& 99\%	& $-7.3_{-5.6}^{+33.9}$ 	& -7.3 \ \cite{wimmer2019discovery}&&2.142	& 0\% 		&                \\
$^{67}$Kr		&		&0.321	& 8\%	&                       			&                       			&		&		&			&                                 \\
$^{68}$Kr  	&		&1.312	& 0\%	&                       			&                       			&		&-0.117	& 67\% 		&                   \\
\textbf{$^{69}$Kr}&		&1.252	& 0\%	&                       			&                       			&		&0.767	& 0\% 		&                    \\
\textbf{$^{70}$Kr}&		&2.465	& 0\%	&                      			&                      			&		&2.016	& 0\%		&                    \\
$^{69}$Rb  	& 3		&-2.856	& 100\%	& $-16.9_{-0.7}^{+0.9}$ 	&                       			& 2		&-1.468	& 100\%		& $2.7_{-5.2}^{+9.1}$      \\
$^{70}$Rb  	& 1		&-1.859	& 100\% 	& $-15.5_{-1.3}^{+1.8}$ 	&                       			& 2		&-0.539	& 98\% 		& $26.6_{-17.5}^{+319.2}$  \\
$^{71}$Rb$^{\rm \#}$ & 3	&-1.479	& 100\% 	& $-13.1_{-1.7}^{+2.7}$ 	&                       			& 		&1.047	& 0\% 		&                          \\
$^{72}$Rb	& 1		&-0.491	& 98\%	& $-5.2_{-6.6}^{+70.1}$ 	& $-7.0_{-0.1}^{+0.08}$\ \cite{suzuki2017discovery}&                 &2.038	& 0\%                &                          \\
$^{73}$Rb	& 1 		&-0.285	& 89\% 	& $2.1_{-11.7}$           	& \textless{}-7.1\ \cite{suzuki2017discovery}&	&4.695	& 0\%                &                         \\
$^{70}$Sr  	&		&0.290	& 10\%	&                       			&                       			& 0		&-2.477	& 100\%              	& $-5.7_{-2.6}^{+3.7}$     \\
$^{71}$Sr  	&		&0.294	& 10\%	&                       			&                       			& 0		&-1.485	& 100\%              	& $2.8_{-5.3}^{+9.1}$      \\
$^{72}$Sr  	&		&1.115	& 0\% 	&                       			&                       			& 0		&-0.290	& 87\%               	& $50.2_{-34.3}$           \\
$^{73}$Sr  	&		&0.975	& 0\% 	&                       			&                       			&		&0.550	& 2\%                		&                          \\
\textbf{$^{74}$Sr}  & 	&1.624	& 0\% 	&                      			&                      			& 		&1.397	& 0\%                		&                          \\
$^{73}$Y   	& 4 		&-2.712	& 100\% 	& $-14.9_{-0.8}^{+1.0}$	&                       			& 3 		&-1.511	& 100\%              	& $4.4_{-5.3}^{+9.1}$      \\
$^{74}$Y   	& 2		&-1.771	& 100\%	& $-15.0_{-1.4}^{+2.1}$	&                       			& 1 		&-0.717	& 100\%              	& $20.7_{-13.3}^{+51.7}$   \\
$^{75}$Y   	& 2		&-1.698	& 100\%	& $-14.7_{-1.5}^{+2.3}$	&                       			&		&-0.003	& 51\%               	&                          \\
$^{76}$Y$^{\rm \#}$	&0	&-0.752	& 100\%  	& $-9.2_{-4.2}^{+12.0}$	&                       			&		&1.613	& 0\%                		&                          \\
$^{77}$Y   	& 2 		&-0.520	& 99\% 	& $-4.1_{-6.5}^{+46.1}$	&                       			&		&3.568	& 0\%                		&                          \\
\textbf{$^{78}$Y}  & 	&1.653	& 0\%	&                       			&                       			&		&5.980	& 0\%                		&                          \\
$^{74}$Zr  	& 		&0.033	& 44\%	&                      			&                       			& 0 		&-2.580	& 100\%              	& $-5.1_{-2.6}^{+3.6}$     \\
$^{75}$Zr  	&		&-0.014	& 52\%	&                       			&                       			& 2		&-1.693	& 100\%              	& $2.5_{-4.7}^{+7.6}$      \\
$^{76}$Zr  	& 		&0.825	& 0\% 	&                       			&                       			& 0 		&-0.790	& 100\%              	& $19.2_{-12.2}^{+39.6}$  \\
$^{77}$Zr  	& 		&0.840	& 0\% 	&                       			&                      			&		&0.165	& 26\%               	&                          \\
\textbf{$^{78}$Zr}   &	&1.786	& 0\%	&                       			&                       			&		&1.335	& 0\%                		&                          \\
\textbf{$^{79}$Zr}   &	&1.810	& 0\%	&                      			&                       			& 		&3.525	& 0\%                		&                          \\
\textbf{$^{80}$Zr}   &	&3.887	& 0\%	&                       			&                       			&		&5.634	& 0\%                		&                          \\
$^{77}$Nb  	& 2  		&-2.720	& 100\% 	& $-17.2_{-0.8}^{+1.1}$	&                       			& 0		&-1.798	& 100\%              	& $1.6_{-4.4}^{+7.0}$        \\
$^{78}$Nb  	& 1  		&-1.878	& 100\% 	& $-15.6_{-1.4}^{+2.0}$	&                       			& 1		&-0.948	& 100\%              	& $15.7_{-10.0}^{+25.4}$    \\
$^{79}$Nb  	& 1 		&-1.643	& 100\% 	& $-14.7_{-1.6}^{+2.5}$	&                       			&		&0.225	& 20\%               	&                          \\
$^{80}$Nb  	& 1 		&-0.332	& 93\%  	& $2.5_{-11.1}$         	&                       			& 		&1.553	& 0\%                		&                          \\
$^{81}$Nb  	& 1 		&-0.480	& 98\% 	& $-2.8_{-7.5}^{+101.3}$  & \textless{}-7.4\ \cite{suzuki2017discoveryMo}&&3.475	& 0\%                &                          \\
\textbf{$^{82}$Nb}	&	&1.336	& 0\%	&                       			&                       			&		&5.150	& 0\%                		&                         \\
$^{78}$Mo  	& 2  		&-0.372	& 95\% 	& $2.19_{-10.1}$		&                       			& 0 		&-2.982	& 100\%              	& $-6.1_{-2.3}^{+3.0}$       \\
$^{79}$Mo  	& 2  		&-0.330	& 92\% 	& $4.1_{-11.4}$		&                       			& 1 		&-2.105	& 100\%              	& $-0.3_{-3.7}^{+5.4}$    \\
$^{80}$Mo  	& 		&0.489	& 2\%	&                       			&                       			& 0 		&-1.059	& 100\%              	& $13.9_{-9.1}^{+20.2}$    \\
$^{81}$Mo  	&		&0.284	& 11\% 	&                       			&                       			& 		&0.040	& 44\%               	&                         \\
\textbf{$^{82}$Mo}	& 	&1.469	& 0\%  	&                       			&                      			&		&1.069	& 0\%                		&                          \\
\textbf{$^{83}$Mo}	& 	&1.824	& 0\%	&                       			&                       			& 		&3.234	& 0\%                		&                          \\
\textbf{$^{84}$Mo}	&	&3.360	& 0\%	&                       			&                       			& 		&5.131	& 0\%                		&                          \\
$^{81}$Tc  	& 1		&-3.059	& 100\% 	& $-18.2_{-0.7}^{+0.9}$	&                       			& 2 		&-2.462	& 100\%              	& $-2.0_{-3.1}^{+4.2}$       \\
$^{82}$Tc  	& 2 		&-2.246	& 100\%	& $-15.7_{-1.1}^{+1.5}$	&                       			& 3		&-1.861	& 100\%              	& $3.5_{-4.5}^{+6.9}$      \\
$^{83}$Tc  	& 1		&-1.936	& 100\% 	& $-15.4_{-1.4}^{+2.0}$ 	&                       			& 2 		&-0.373	& 92\%               	& $50.1_{-30.4}$           \\
$^{84}$Tc$^{\rm \#}$&2 	&-1.031	& 100\% 	& $-9.7_{-3.2}^{+6.5}$	&                       			&		&0.879	& 0\%                		&                          \\
$^{85}$Tc$^{\rm \#}$&1	&-0.724	& 100\% 	& $-6.7_{-4.9}^{+14.7}$	& \textless{}-7.4\ \cite{suzuki2017discoveryMo}	&	&2.714	& 0\%                &                          \\
\textbf{$^{86}$Tc}   & 	&0.863	& 0\% 	&                      			&                       			&		&4.616	& 0\%                		&                          \\
$^{82}$Ru  	& 1		&-0.429	& 97\%	& $0.6_{-9.1}$          		&                       			& 0		&-3.367	& 100\%              	& $-6.8_{-2.1}^{+2.6}$    \\
$^{83}$Ru  	& 4 		&-0.679	& 100\%	& $-2.5_{-5.4}^{+18.1}$	&                       			& 2		&-2.812	& 100\%              	& $-3.6_{-2.6}^{+3.5}$    \\
$^{84}$Ru  	&  		&0.244	& 14\% 	&                       			&                       			& 0 		&-1.586	& 100\%              	& $6.5_{-5.7}^{+9.5}$     \\
$^{85}$Ru  	&		&0.179	& 22\% 	&                       			&                       			& 2		&-0.754	& 100\%              	& $26.1_{-14.4}^{+51}$    \\
$^{86}$Ru  	&		&0.996	& 0\%	&                       			&                       			&		&0.364	& 8\%                		&                         \\
\textbf{$^{87}$Ru}	& 	&1.138	& 0\% 	&                       			&                      			& 		&2.085	& 0\%                		&                         \\
\textbf{$^{88}$Ru}	&	&3.276	& 0\%	&                       			&                       			&		&4.372	& 0\%                		&                         \\
\textbf{$^{89}$Ru}	& 	&3.509	& 0\%	&                       			&                       			& 		&5.613	& 0\%                		&                         \\
$^{85}$Rh  	& 4 		&-3.318	& 100\%	& $-15.3_{-0.7}^{+0.9}$	&                       			& 3 		&-2.955	& 100\%              	& $-3.3_{-2.5}^{+3.3}$    \\
$^{86}$Rh  	& 2 		&-2.603	& 100\% 	& $-16.3_{-1.0}^{+1.3}$	&                       			& 6 		&-2.313	& 100\%              	& $3.3_{-3.5}^{+5.0}$       \\
$^{87}$Rh  	& 4 		&-2.427	& 100\%	 & $-13.5_{-1.1}^{+1.4}$	&                       			& 3 		&-1.327	& 100\%              	& $12.3_{-7.3}^{+13.8}$   \\
$^{88}$Rh  	& 0 		&-1.441	& 100\%	 & $-13.0_{-2.2}^{+3.5}$	&                       			& 6 		&-0.206	& 78\%               	& $87.9_{-56.7}$          \\
$^{89}$Rh$^{\rm \#}$&4	&-1.223	& 100\%	& $-4.6_{-2.7}^{+4.8}$	& \textless{}-6.9\ \cite{vcelikovic2016new}&&2.143	& 0\%                &                         \\
\textbf{$^{90}$Rh}	& 	&1.563	& 0\%	&                       			&                       			& 		&5.154	& 0\%                		&                         \\
\textbf{$^{91}$Rh}	&	&0.851	& 0\%	&                       			&                       			&  		&5.716	& 0\%                		&                         \\
$^{86}$Pd  	& 4 		&-0.259	& 87\% 	& $13.7_{-16.1}$		&                       			& 0		&-3.445	& 100\%              	& $-6.2_{-2.1}^{+2.7}$    \\
$^{87}$Pd  	& 5		&-0.462	& 98\%	& $5.3_{-8.8}$          		&                       			& 5		&-2.940	& 100\%              	& $-1.2_{-2.6}^{+3.4}$    \\
$^{88}$Pd$^{\rm \dagger}$& &0.568	& 1\%	&                       			&                       			& 0 		&-1.742	& 100\%              	& $6.1_{-5.3}^{+8.4}$     \\
$^{89}$Pd  	&		&0.444	& 3\% 	&                       			&                       			& 0 		&-0.887	& 100\%              	& $23.1_{-12.4}^{+34.4}$  \\
$^{90}$Pd  	&		&1.696	& 0\%	&                       			&                       			& 		&0.575	& 1\%                		&                         \\
\textbf{$^{91}$Pd}  & 	&0.970	& 0\% 	&                       			&                       			&		&2.628	& 0\%                		&                         \\
\textbf{$^{92}$Pd}  & 	&3.587	& 0\% 	&                       			&                       			& 		&4.526	& 0\%                		&                         \\
\textbf{$^{93}$Pd}  &	&3.507	& 0\% 	&                       			&                       			&		&5.636	& 0\%                		&                         \\
$^{89}$Ag  	& 4 		&-3.143	& 100\% 	& $-14.8_{-0.8}^{+1.0}$	&                       			& 2		&-2.445	& 100\%              	& $0.6_{-3.5}^{+4.8}$     \\
$^{90}$Ag  	& 4 		&-2.183	& 100\%	& $-12.4_{-1.3}^{+1.8}$	&                       			& 4 		&-1.616	& 100\%              	& $10.1_{-5.9}^{+10}$     \\
$^{91}$Ag  	& 4  		&-2.192	& 100\% 	& $-12.5_{-1.3}^{+1.8}$	&                       			& 0 		&-0.381	& 92\%               	& $56.3_{-32.9}$          \\
$^{92}$Ag  	& 2  		&-1.024	& 100\%	& $-8.2_{-3.5}^{+7.2}$	&                       			&		&0.053	& 42\%               	&                         \\
$^{93}$Ag$^{\rm \#}$&4 	&-1.119	& 100\% 	& $-6.9_{-3.1}^{+6.0}$	& $-6.6_{-0.03}^{+0.03}$\ \cite{vcelikovic2016new}  &&2.568	& 0\%                &                         \\
\textbf{$^{94}$Ag}	& 	&1.423	& 0\%	&                       			&                       			& 		&5.023	& 0\%                		&                         \\
\textbf{$^{95}$Ag}	& 	&0.958	& 0\%	&                       			&                       			&		&5.598	& 0\%                		&                         \\
$^{90}$Cd  	& 4 		&-0.179	& 78\% 	& $23.6_{-23.8}$		&                       			& 0 		&-3.180	& 100\%              	& $-3.9_{-2.4}^{+3.2}$    \\
$^{91}$Cd  	& 4 		&-0.147	& 74\% 	& $28.8_{-28.3}$		&                       			& 0 		&-2.195	& 100\%              	& $2.8_{-4.1}^{+5.9}$     \\
$^{92}$Cd$^{\rm \dagger}$& &0.612	& 0\% 	&                       			&                       			& 0		&-1.453	& 100\%              	& $11.9_{-7.1}^{+12.5}$   \\
$^{93}$Cd  	&		&0.537	& 1\%	&                       			&                       			& 0 		&-0.367	& 92\%               	& $60.1_{-35.1}$          \\
$^{94}$Cd  	&		&1.556	& 0\%	&                       			&                       			& 		&0.550	& 2\%                		&                         \\
\textbf{$^{95}$Cd}	&	&1.076	& 0\%  	&                       			&                       			&		&2.605	& 0\%                		&                         \\
\textbf{$^{96}$Cd}	&	&3.439	& 0\% 	&                       			&                       			&		&4.495	& 0\%                		&                         \\
\textbf{$^{97}$Cd}	&	&3.317	& 0\% 	&                       			&                       			&		&5.545	& 0\%                		&                         \\
$^{93}$In  	& 4 		&-3.041	& 100\% 	& $-14.3_{-0.9}^{+1.1}$	&                       			& 0		&-2.288	& 100\%              	& $2.7_{-4.0}^{+5.6}$       \\
$^{94}$In  	& 0 		&-2.105	& 100\%	& $-15.0_{-1.4}^{+2.0}$	&                       			& 2		&-1.435	& 100\%              	& $13.5_{-7.3}^{+13.2}$   \\
$^{95}$In  	& 4 		&-2.215	& 100\%	& $-12.2_{-1.3}^{+1.8}$	&                       			& 0 		&-0.533	& 98\%               	& $45.1_{-24.1}$          \\
$^{96}$In  	& 4		&-1.065	& 100\%	 & $-5.8_{-3.5}^{+7.0}$	&                       			&		&0.130	& 31\%               	&                         \\
$^{97}$In$^{\rm \#}$  &4	&-1.201	& 100\% 	& $-7.0_{-3.0}^{+5.5}$	&					&		&2.349	& 0\%                		&                         \\
\textbf{$^{98}$In}  &	&1.072	& 0\% 	&                       			&                       			&		&4.494	& 0\%                		&                         \\
\textbf{$^{99}$In}  &	&1.207	& 0\%	&                       			&                       			&		&5.657	& 0\%                		&                         \\
$^{94}$Sn  	& 4 		&-0.410	& 96\% 	& $7.8_{-11}$			&                       			& 0 		&-3.297	& 100\%              	& $-3.4_{-2.5}^{+3.1}$    \\
$^{95}$Sn  	& 0 		&-0.462	& 98\%	& $2.6_{-9.6}$          		&                       			& 0		&-2.421	& 100\%              	& $2.2_{-3.8}^{+5.3}$     \\
$^{96}$Sn  	&		&0.291	& 11\% 	&                       			&                       			& 0 		&-1.785	& 100\%              	& $8.7_{-5.7}^{+8.9}$     \\
$^{97}$Sn  	&		&0.199	& 20\%	&                       			&                       			& 0 		&-0.735	& 100\%              	& $34.2_{-17.1}^{+66.4}$  \\
$^{98}$Sn  	&		&1.205	& 0\% 	&                       			&                       			&  		&0.128	& 32\%               	&                         \\
\textbf{$^{99}$Sn}   &	&1.326	& 0\% 	&                      			&                       			& 		&2.515	& 0\%                		&                         \\
$^{100}$Sn 	&		&3.474	& 0\% 	&                       			&                       			&		&		&          			&                    \\

%\end{tabular*}
%\end{center}
\end{longtable*}

\section{Conclusion}

	In summary, based on the nuclear shell model, two formulas, with respectively four and three terms, are recommended in this work for calculating the nuclear binding energies with the shell-model binding energies in the $30\leqslant Z, N\leqslant50$ region. The contributions of Coulomb interaction, neutron shell effect and isospin effect are well included. After applying the Bootstrap framework, these two formulas have the total uncertainties of around 0.3 MeV for the nuclides with experimentally known binding energies. The repulsion characteristic of Coulomb interaction decreases the binding energy as expected. It is found that the neutron shell effect and the isospin effect compensate with each other when it is far from the proton dripline.  For those without experimental values, the uncertainty of predicted data is assessed to be around 0.4 MeV, which shows a nice extrapolation power. In addition, the formulas for $S_{p}$ and $S_{2p}$ are also recommended with uncertainties less than 0.3 MeV. It shows that the predicted uncertainties of proton(s) separation energies are mostly smaller than those of the AME2016 extrapolations and the proton dripline can be extended comparing with the boundary reached in experiment. The Bootstrap method is developed to estimate the possibilities of proton(s)-emission $P(S_{p}<0)$ and $P(S_{2p}<0)$ from the distribution of $S_{p}$ and $S_{2p}$. The prediction of proton(s)-emission property of nuclides is mostly consistent with the experimental results. 30 nuclides are suggested to be bound against both $p$-emission and $2p$-emission. Their spectroscopic properties, such as $\beta$-decay spectrum, need to be investigated experimentally and theoretically in future. The energies and partial half-lives predicted in this work can be used as inputs for the simulation of nuclear astrophysics.

\begin{acknowledgments}

	This work has been supported by the National Natural Science Foundation of China under Grant Nos. 11775316, 11825504, 11961141004, the Tip-top Scientific and Technical Innovative Youth Talents of Guangdong special support program under Grant No. 2016TQ03N575, the Guangdong Major Project of Basic and Applied Basic Research under Grant No. 2021B0301030006, the computational resources from SYSU and National Supercomputer Center in Guangzhou.

\end{acknowledgments}

%\bibliography{apstemplate}

\begin{thebibliography}{99}
\bibitem {bertulani2020} C. A. Bertulani, Sci. China-Phys. Mech. Astron., 63(11): 112063 (2020)

\bibitem {liu2020network} H. L. Liu, D. D. Han, Y. G. Ma $et \ al.$, Sci. China-Phys. Mech. Astron., 63(11): 112062 (2020)

\bibitem {kappeler2011s} F. K{\"a}ppeler, R. Gallino, S. Bisterzo $et \ al.$, Rev. Mod. Phys., 83(1): 157 (2011)

\bibitem {arnould2007r} M. Arnould, S. Goriely, and K. Takahashi, Phys. Rep., 450(4-6): 97 (2007)

\bibitem {arnould2003p} M. Arnould and S. Goriely, Phys. Rep., 384(1-2): 1 (2003)

\bibitem {thoennessen2011isotopes} M. Thoennessen and B. Sherrill, Nature, 473(7345): 25 (2011)

\bibitem {erler2012limits} J. Erler, N. Birge, M. Kortelainen $et \ al.$, Nature, 486(7404): 509 (2012)

\bibitem {nazarewicz2018limits} W. Nazarewicz, Nat. Phys., 14(6): 537 (2018)

\bibitem {thoennessen2004reaching} M. Thoennessen, Rep. Prog. Phys., 67(7): 1187 (2004)

\bibitem {thoennessen2016discovery} M. Thoennessen, \textit{The Discovery of Isotopes}, First edition (Switzerland: Springer International Publishing AG Switzerland, 2016) 

\bibitem {ahn2019location} D. Ahn, N. Fukuda, H. Geissel $et \ al.$, Phys. Rev. Lett., 123(21): 212501 (2019)

\bibitem {myers1966nuclear} W. D. Myers and W. J. Swiatecki, Nucl. Phys., 81(1): 1 (1966)

\bibitem {myers1976development} W. D. Myers, At. Data Nucl. Data Tables, 17(5-6): 411 (1976)

\bibitem {moller1981atomic} P. M{\"o}ller and J. R. Nix, At. Data Nucl. Data Tables, 26(2): 165 (1981)

\bibitem {moller1981nuclear} P. M{\"o}ller and J. R. Nix, Nucl. Phys. A, 361(1): 117 (1981)

\bibitem {moller1988nuclear} P. M{\"o}ller, W. Myers, W. Swiatecki $et \ al.$, At. Data Nucl. Data Tables, 39(2): 225 (1988)

\bibitem {moller1988nuclear2} P. M{\"o}ller and J. Nix, At. Data Nucl. Data Tables, 39(2): 213 (1988)

\bibitem {moller1993nuclear} P. M{\"o}ller, J. Nix, W. Myers $et \ al.$, arXiv preprint nucl-th/9308022 (1993)

\bibitem {moller2016nuclear} P. M{\"o}ller, A. J. Sierk, T. Ichikawa $et \ al.$, At. Data Nucl. Data Tables, 109-110: 1 (2016)

\bibitem {goriely2013hartree} S. Goriely, N. Chamel, and J. Pearson, Phys. Rev. C, 88(6): 061302(R) (2013)

\bibitem {wang2014surface} N. Wang, M. Liu, X. Wu $et \ al.$, Phys. Lett. B, 734: 215 (2014)

\bibitem {ge2019effect} Y. H. Ge, Y. Zhang, and J. N. Hu, Sci. China-Phys. Mech. Astron., 63(4): 242011 (2020)

\bibitem {wu2020production} Z. J. Wu and L. Guo, Sci. China-Phys. Mech. Astron., 63(4): 242021 (2020)

\bibitem {schatz2017dependence} H. Schatz and W. J. Ong, Astrophys. J., 844(2): 139 (2017)

\bibitem {parikh2009impact} A. Parikh, J. Jos{\'e}, C. Iliadis $et \ al.$, Phys. Rev. C, 79(4): 045802 (2009)

\bibitem {dobaczewski2014} J. Dobaczewski, W. Nazarewicz, and P. G. Reinhard, J. Phys. G: Nucl. Part. Phys., 41(7): 074001 (2014)

\bibitem {yuan2016uncertainty} C. X. Yuan, Phys. Rev. C, 93(3): 034310 (2016)

\bibitem {cai2020alpha} B. S. Cai, G. S. Chen, J. Y. Xu $et \ al.$, Phys. Rev. C, 101(5): 054304 (2020)

\bibitem {efron1982jackknife} B. Efron, \textit{The jackknife, the bootstrap, and other resampling plans}, First edition (Philadelphia: Society for Industrial and Applied Mathematics, 1982)

\bibitem {efron1992bootstrap} B. Efron, Bootstrap Methods: Another Look at the Jackknife. In \textit{Breakthroughs in statistics Vol. II} (New York: Springer-Verlag New York, Inc., 1992), p. 569

\bibitem {lee2020isospin} J. Lee, X. X. Xu, K. Kaneko $et \ al.$, Phys. Rev. Lett., 125(19): 192503 (2020)

\bibitem {jian2021isospin} H. Jian, Y. F. Gao, F. C. Dai $et \ al.$, Symmetry, 13(12): 2278 (2021)

\bibitem {xu2017Si22} X.X. Xu, C. J. Lin, L. J. Sun $et \ al.$, Phys. Lett. B, 766: 312 (2017)

\bibitem {liang2020P26} P. F. Liang, L. J. Sun, J. Lee $et \ al.$, Phys. Rev. C, 101(2): 024305 (2020)

\bibitem {sun2019S27} L. J. Sun, X. X. Xu, C. J. Lin $et \ al.$, Phys. Rev. C, 99(6): 064312 (2019)

\bibitem {wu2020Al22} C. G. Wu, H. Y. Wu, J. G. Li $et \ al.$, Phys. Rev. C, 104(4): 044311 (2021)

\bibitem {wang2021Na20} Y. B. Wang, J. Su, Z. Y. Han $et \ al.$, Phys. Rev. C, 103(1): L011301 (2021)

\bibitem {jin2021Mg18} Y. Jin, C. Y. Niu, K. W. Brown $et \ al.$, Phys. Rev. Lett., 127(26): 262502 (2021)

\bibitem {zhang2021U214} Z. Y. Zhang, H. B. Yang, M. H. Huang $et \ al.$, Phys. Rev. Lett., 126(15): 152502 (2021)

\bibitem {zhang2020Pa218} M. M. Zhang, H. B. Yang, Z. G. Gan $et \ al.$, Phys. Lett. B, 800: 135102 (2020) 

\bibitem {zhou2021Th213} H. B. Zhou, Z. G. Gan, N. Wang $et \ al.$,  Phys. Rev. C, 103(4): 044314 (2021)

\bibitem {yuan2016Sn132} C. X. Yuan, Z. Liu, F. R. Xu $et \ al.$, Phys. Lett. B, 762: 237 (2016) 

\bibitem {chen2019Sn132}Z. Q. Chen, Z. H. Li, H. Hua $et \ al.$, Phys. Rev. Lett., 122(21), 212502 (2019)

\bibitem {watanabe2021Ag127} H. Watanabe, C. X. Yuan, G. Lorusso $et \ al.$,  Phys. Lett. B, 823: 136766 (2021)

\bibitem {yun2020what} X. Y. Yun, D. Y. Pang, Y. P. Xu $et \ al.$, Sci. China-Phys. Mech. Astron., 63(2): 222011 (2020)

\bibitem {xu2019In101} X. Xu, J. H. Liu, C. X. Yuan $et \ al.$, Phys. Rev. C, 100(5): 051303(R) (2019)

\bibitem {jin2021Cd98} S. Y. Jin, S. T. Wang, J. Lee $et \ al.$, Phys. Rev C, 104(2): 024302 (2021)

\bibitem {honma2009new} M. Honma, T. Otsuka, T. Mizusaki $et \ al.$, Phys. Rev. C, 80(6): 064323 (2009)

\bibitem {herndl1997shell} H. Herndl and B. Brown, Nucl. Phys. A, 627(1): 35 (1997)

\bibitem {caurier1999full} E. Caurier, G. Martinez-Pinedo, F. Nowacki $et \ al.$, Phys. Rev. C, 59(4): 2033 (1999)

\bibitem {jia2021possible} J. Jia, Y. Qian, and Z. Ren, Phys. Rev. C, 104(3): L031301 (2021)

\bibitem {wang2017ame2016} M. Wang, G. Audi, F. Kondev $et \ al.$, Chin. Phys. C, 41(3): 030003 (2017)

\bibitem {huang2017ame2016} W. J. Huang, G. Audi, M. Wang, $et \ al.$, Chin. Phys. C, 41(3), 030002 (2017)

\bibitem {brown2014shell} B. Brown and W. Rae, Nucl. Data Sheets, 120: 115 (2014)

\bibitem {pruet2006nucleosynthesis} J. Pruet, R. D. Hoffman, S. E. Woosley $et \ al.$, Astrophys. J., 644(2): 1028 (2006)

\bibitem {qi2012effects} C. Qi, D. S. Delion, R. J. Liotta $et \ al.$, Phys. Rev. C, 85(1): 011303(R) (2012)

\bibitem {hoff2020mirror} D. E. M. Hoff, A. M. Rogers, S. M. Wang $et \ al.$, Nature, 580(7801): 52 (2020)



\bibitem {vcelikovic2016new} I. {\'C}elikovi{\'c}, M. Lewitowicz, R. Gernh{\"a}user $et \ al.$, Phys. Rev. Lett., 116(16): 162501 (2016)

\bibitem {suzuki2017discovery} H. Suzuki, L. Sinclair, P.-A. S{\"o}derstr{\"o}m $et \ al.$, Phys. Rev. Lett., 119(19): 192503 (2017)

\bibitem {sinclair2019half} L. Sinclair, R. Wadsworth, J. Dobaczewski $et \ al.$, Phys. Rev. C, 100(4): 044311 (2019)

\bibitem {kaneko2013variation} K. Kaneko, Y. Sun, T. Mizusaki $et \ al.$, Phys. Rev. Lett., 110(17): 172505 (2013)

\bibitem {wimmer2019discovery} K. Wimmer, P. Doornenbal, W. Korten $et \ al.$, Phys. Lett. B, 795: 266 (2019)

\bibitem {suzuki2017discoveryMo} H. Suzuki, T. Kubo, N. Fukuda $et \ al.$, Phys. Rev. C, 96(3): 034604 (2017)

%\bibitem {moller2019nuclear} P. M{\"o}ller, M. R. Mumpower, T. Kawano $et \ al.$, At. Data Nucl. Data Tables., 125: 1 (2019)

%\bibitem {woosley1978p} S. E. Woosley and W. M. Howard, Astrophys. J. Suppl. Ser., 36(2): 285 (1978)

%\bibitem {rayet1990p} M. Rayet, M. Arnould, and N. Prantzos, Astron. Astrophys., 227: 271 (1990)

%\bibitem {howard1993} W. M. Howard and B. S. Meyer, in \textit{Proceedings of the 2nd International Symposium on Nuclear Astrophysics}, edited by F. K{\"a}ppeler and K. Wisshak (Bristol: IOP Publishing, 1993), p. 575

%\bibitem {goriely2005p} S. Goriely, D. Garc{\'\i}a-Senz, E. Bravo $et \ al.$, Astron. Astrophys., 444(1): L1 (2005)

%\bibitem {travaglio2011type} C. Travaglio, F. K. R{\"o}pke, R. Gallino $et \ al.$, Astrophys. J., 739(2): 93 (2011)

%\bibitem {schatz1998rp} H. Schatz, A. Aprahamian, J. G{\"o}rres $et \ al.$, Phys. Rep., 294(4): 167 (1998)

%\bibitem {frohlich2006neutrino} C. Fr{\"o}hlich, G. Mart{\'\i}nez-Pinedo, M. Liebend{\"o}rfer $et \ al.$, Phys. Rev. Lett., 96(14): 142502 (2006)



%\bibitem {wanajo2011uncertainties} S. Wanajo, H.-T. Janka, and S. Kubono, Astrophys. J., 729(1): 46 (2011)

%\bibitem {arcones2012impact} A. Arcones, C. Fr{\"o}hlich, and G. Mart{\'\i}nez-Pinedo, Astrophys. J., 750(1): 18 (2012)


\end{thebibliography}

\end{document}